\documentclass{article}
\usepackage[a4paper, total={6in, 10in}]{geometry}
\usepackage{indentfirst}
\usepackage{bbm}
\usepackage{physics}
\usepackage{mdframed}
\usepackage{float}
\usepackage[caption = false]{subfig}
\usepackage{graphicx,color}
\usepackage{amsbsy,amssymb,amsmath,bm,mathrsfs,mathdots}
\usepackage{cite}
\usepackage{authblk}

\usepackage{tabularray}
\usepackage{makecell}
\usepackage[title]{appendix}
\apptocmd{\appendices}{\apptocmd{\thesection}{}{}{}}{}{}
\usepackage{hyperref}
\hypersetup{
    colorlinks=true,
    linkcolor=blue,
    filecolor=blue,      
    urlcolor=blue,
    citecolor=blue,   
    pdftitle={Overleaf Example},
    pdfpagemode=FullScreen,
    }

\def \bea {\begin{eqnarray}}
\def \eea {\end{eqnarray}}

\makeatother

\usepackage{amscd}% for diagram. added by ken

\begin{document}
\title{Connection between Free-Fermion and Interacting Crystalline Symmetry-Protected Topological Phases}

\author[1,$\ast$]{Chen-Shen Lee}
\author[2,$\dagger$]{Ken Shiozaki}
\author[1,3,4,$\ddagger$]{Chang-Tse Hsieh}
\affil[1]{Department of Physics and Center for Theoretical Physics, National Taiwan University,
Taipei 10607, Taiwan}
\affil[2]{Center for Gravitational Physics and Quantum Information, Yukawa Institute for Theoretical Physics, Kyoto University, Kyoto 606-8502, Japan}
\affil[3]{Physics Division, National Center for Theoretical Science, National Taiwan University,
Taipei 10607, Taiwan}
\affil[4]{Center for Quantum Science and Engineering, National Taiwan University, Taipei 10617, Taiwan}
\affil[$\ast$]{yausan0523270@gmail.com}
\affil[$\dagger$]{ken.shiozaki@yukawa.kyoto-u.ac.jp}
\affil[$\ddagger$]{cthsieh@phys.ntu.edu.tw}
\date{\today }

\maketitle
\selectfont
We present a framework for investigating the effects of interactions on crystalline symmetry-protected topological (SPT) phases. Within this framework, one can establish a direct connection between the equivalence classes of free-fermion systems and their corresponding interacting classes. A central component of this framework is the Atiyah-Hirzebruch spectral sequence, which provides a systematic way to represent crystalline SPT phases as SPT phases with internal symmetries on subspaces. We demonstrate the application of this approach through examples in various dimensions: $1d$ systems with $U(1)$ and reflection symmetry, $2d$ systems with $U(1)$ and $C_n$ rotation symmetry, and $3d$ systems with $U(1)$ and inversion symmetry.

%We propose a prescription for investigating the interaction effects on crystalline symmetry-protected topological (SPT) phases. Specifically, using this method, one can establish a clear connection between the equivalence class a free-fermion system falls into and its corresponding interacting class. A key element of this prescription is the Atiyah-Hirzebruch spectral sequence (AHSS), which offers a systematic method to represent crystalline SPT phases as (on-site) SPT phases on subspaces. In this work, we demonstrate the application of this approach through three examples across different dimensions: $1d$ systems with $U(1)$ and reflection symmetry, $2d$ systems with $U(1)$ and $C_n$ rotation symmetry, and $3d$ systems with $U(1)$ and inversion symmetry.

\section{Introduction}
In recent years, there has been rapid progress in topological quantum matter, stemming from the foundational work on the integer quantum Hall effect~\cite{IQHE} and further driven by the discovery of the quantum spin Hall effect~\cite{IQSHE1,IQSHE2,IQSHE3,IQSHE4} and strong topological insulators~\cite{TI1,TI2}. These latter discoveries have sparked a wide range of research into free-fermion topological systems. Such systems can be classified for all 10 Altland-Zirnbauer (AZ) symmetry classes~\cite{AZclass} using K-theory~\cite{Ktheory1,Ktheory2,Ktheory3,Ktheory4}. Beyond the Chern class (Class A), systems in other classes are constrained by internal symmetries, and thus their quantum phases are part of the family of symmetry-protected topological (SPT) phases~\cite{SPT1,SPT2}. SPT phases are short-range entangled gapped states that are protected by symmetry. More precisely, distinct SPT states cannot be adiabatically deformed into one another without closing the energy gap while preserving the protecting symmetry. A hallmark of free-fermion SPT phases is the bulk-boundary correspondence, in which non-trivial topological phases inherently lead to the presence of gapless surface states~\cite{BBC1,BBC2,BBC3,BBC4,BBC5,BBC6,BBC7,BBC8,BBC9,BBC10}. Notable examples include the quantum spin Hall system HgTe and strong topological insulators.

Some topological or SPT phases remain robust even in the presence of interactions because their topological properties are tied to physically measurable quantities. For example, 2$d$ Chern insulators are characterized by the Chern number, which corresponds to the Hall conductivity. Similarly, 3$d$ topological insulators protected by time-reversal symmetry have a $\mathbb{Z}_2$ classification related to magneto-electric polarization~\cite{3DTIwithTR}. On the other hand, certain phases are reduced by interactions~\cite{edgedegeneracy1,edgedegeneracy2,reductionSPT1,reductionSPT2,reductionSPT3}, as some distinct free-fermion phases can be adiabatically connected via strong interactions. One way to study the interaction effects is through bulk-boundary correspondence. For instance, a 1$d$ time-reversal topological superconductor formed by stacking 8 non-trivial Kitaev chains becomes topologically trivial in the many-body regime~\cite{edgedegeneracy1}. This illustrates the connection between the free-fermion classification $\mathbb{Z}$ and the interacting classification $\mathbb{Z}_8$. Another example involves using edge degeneracy to investigate interaction effects~\cite{edgedegeneracy2}.

The concept of SPT phases can be generalized to crystalline SPT phases by extending the protecting symmetry to crystalline symmetry (see~\cite{TCI1,TCI2,TCI3,TCI4,TCI5,TCI6,TCI7} for free-fermion systems and~\cite{CSPT1,CSPT2,CSPT3,CSPT4,CSPT5,CSPT6,CSPT7} for interacting systems). However, unlike SPT phases, bulk-boundary correspondence in free-fermion crystalline SPT phases does not always ensure the presence of gapless surface states, making it harder to study the interaction effect using bulk-boundary correspondence, leaving this area less understood. The interaction effects have been studied only in a few rare cases where the connection between free-fermion and many-body topological invariants is clear, such as in 2$d$ systems with $U(1)$, discrete magnetic translation, and $M$-fold point group rotation symmetries ($M=1,2,3,4,6$) under the condition of zero magnetic flux per unit cell.~\cite{free-to-interactingmap}. Here, we propose a systematic approach to studying the interaction effects on free-fermion crystalline SPT phases. The central idea of this approach is using the Atiyah-Hirzebruch spectral sequence (AHSS)~\cite{AHSSsource}, a powerful mathematical tool used to compute generalized (co)homology, to construct crystalline SPT phases. Specifically, crystalline SPT phases can be built from lower-dimensional SPT states~\cite{constructionCSPTfromlowerd1,constructionCSPTfromlowerd2,Lower-dconstruction1,Lower-dconstruction2,Lower-dconstruction3,2024realspaceconstruction}, and the AHSS provides a structured method for implementing this lower-dimensional construction in both interacting~\cite{interactingAHSS} and free-fermion contexts~\cite{freeAHSSinkspace,freeAHSSinrealspace}. Through the AHSS, crystalline SPT phases can be expressed in terms of SPT phases (protected by on-site symmetries) on subspaces. With this SPT-phase expression, we establish a clear connection between free-fermion and interacting crystalline SPT phases, providing a better understanding of interaction effects here.

This paper is organized as follows. In Sec.~\ref{Overview of the AHSS}, we briefly introduce the framework of the Atiyah-Hirzebruch spectral sequence (AHSS), showing that the classification of crystalline SPT phases fits into a series of short exact sequences associated with limiting pages. These limiting pages represent SPT phases on a specific subspace that can trivially extend to lower-dimensional subspaces and cannot be trivialized by the SPT states in higher-dimensional subspaces. In other words, the crystalline SPT phases can be constructed from SPT phases on subspaces by solving a series of group extension problems. In the remainder of this paper, we show that limiting pages can be characterized by certain quantum numbers, allowing interaction effects to be analyzed through exploring the relation between quantum numbers in free-fermion and interacting settings. To illustrate that our approach accommodates both split and non-split short exact sequences, we apply it to analyze interaction effects on crystalline SPT phases for 1$d$ systems with $U(1)$ and reflection symmetry and for 2$d$ systems with $U(1)$ and $C_n$ rotation symmetry in Sec.~\ref{Main reflection Sec} and Sec.~\ref{Main Cn rotation sec}, respectively, as examples of split cases. We then explore 3$d$ systems with $U(1)$ and inversion symmetry in Sec.~\ref{Main inversion Sec}, as an example of non-split case. 

\section{Overview of the Atiyah-Hirzebruch spectral sequence}\label{Overview of the AHSS}
In this section, we briefly introduce how to compute the classification of crystalline SPT phases using the AHSS, with an emphasis on its physical significance. A more detailed description of this methodology can be found in this work~\cite{interactingAHSS}.
\subsection{Cell decomposition and the AHSS}\label{Cell decomposition and the AHSS}
The generalized homology $h_{n}^{G}(X,Y)$ can be used to classify various topological crystalline phenomena in the interacting scope and free fermions as well, where $G$ denotes a crystalline symmetry, and $X$ and $Y$ represent the real space manifolds with $Y \subset X$. The integer $n \in \mathbb{Z}$ indicates the degree of SPT phenomena, with $n = 0$ specifically corresponding to the classification of crystalline SPT phases. The AHSS provides a structured framework for computing $h_{n}^{G}(X, Y)$. This process begins with a $G$-symmetric cell decomposition of the space manifold $X$, followed by defining the $p$-skeleton $X_p$ of $X$
\bea\label{skeleton}
X_0=\{\text{0-cells}\}, \qquad X_p=X_{p-1}\cup\{\text{$p$-cells}\}.
\eea
$p$-cells are $p$-dimensional open cells composing the cell decomposition. By doing so, we can obtain a $G$-symmetric filtration of $X$
\bea\label{skeleton subset}
X_0\subset X_1\subset \ldots\subset X_d=X,
\eea
where $d$ is the dimension of $X$. With the $G$-symmetric cell decomposition, one can embed $p-n$-dimensional SPT phases on $X_p$ into $X$, leading to the filtration of $h_{n}^{G}(X,Y)$
\bea\label{filtration}
F_ph_n:=\text{Im}[h_{n}^{G}(X_p,X_p\cap Y)\rightarrow h_{n}^{G}(X,Y)],
\eea
and
\bea\label{filtration subset}
0\subset F_0h_n\subset \ldots\subset F_dh_n=h_{n}^{G}(X,Y).
\eea
To proceed, we introduce the concept of limiting pages $E_{p,n-p}^{\infty}$. $E^{\infty}_{p,n-p}$ represents a set of $(p-n)$ dimensional SPT phases on $p$-cells that create no anomaly on any adjacent low-dimensional cells and cannot be trivialized by any adjacent high-dimensional cells, namely that $E_{p,n-p}^{\infty}$ can be written as the quotient
\bea\label{Def of limiting page}
E_{p,n-p}^{\infty}\cong F_ph_n/F_{p-1}h_n.
\eea
With eq.~\eqref{filtration subset} and eq.~\eqref{Def of limiting page}, we obtain a series of short exact sequences:
\bea\label{short exact sequences}
\begin{alignedat}{3}
0\rightarrow & \ \ \, E^{\infty}_{0,n}&&\rightarrow \quad F_{1}h_{n}&&\rightarrow E^{\infty}_{1,n-1}\rightarrow 0,\\
0\rightarrow & \ \ \, F_{1}h_{n}&&\rightarrow \quad F_{2}h_{n}&&\rightarrow E^{\infty}_{2,n-2}\rightarrow 0,\\
&\vdots\\
0\rightarrow &F_{d-1}h_{n}&&\rightarrow h_{n}^{G}(X,Y)&&\rightarrow E^{\infty}_{d,n-d}\rightarrow 0.\\
\end{alignedat}
\eea
Utilizing the above sequences and given $E^{\infty}$, the generalized homology $h_{n}^{G}(X,Y)$ and each $F_p h_n$ can be computed iteratively. The remaining part is determining the $E^{\infty}$-page, which also can be done in the framework of the AHSS.

Let's start with the $E^{1}$-page of the AHSS. For each $p$-cell $D^{p}_j$, there is a little group $G_{D^{p}_j} \subset G$ that acts on $D^{p}_j$ as on-site symmetry, implying that the $G$-symmetric cell decomposition results in the "local data of SPT phases on $D^{p}_j$", $h_{p-q}^{G_{D^p_j}}(D^{p}_j,\partial D^{p}_j)$. The collection of these local data defines the $E^1$-page, given by
\bea\label{E1-page}
E^1_{p,-q}=\prod _j h_{p-q}^{G_{D^{p}_j}}(D^{p}_j,\partial D^{p}_j),
\eea
where $q$ is the dimension of local SPT phases, and $j$ runs over the set of inequivalent $p$-cells of $X$. The bulk-boundary correspondence offers another interpretation of the $E^1$-page: $E^1_{p,-q}$ represents anomalies over $(q-1)$-dimensional real space with onsite $G_{D^p_j}$ symmetry. To properly glue these local data of SPT phases together, we use the first differential:
\bea\label{1-differential}
d^{1}_{p,-q}:E^{1}_{p,-q}\rightarrow E^{1}_{p-1,-q}.
\eea
The first differential can be regarded as pumping $q$-dimensional SPT states in $p$-cells onto $(p-1)$-cells. Because the boundary of the boundary is trivial, $d^{1}_{p,-q}\circ d^{1}_{p+1,-q}=0$, one can take the homology of $d^1$ and define the $E^2$-page
\bea\label{E2-page}
E^{2}_{p,-q}:=\text{Ker}(d^{1}_{p,-q})/\text{Im}(d^{1}_{p+1,-q}).
\eea
With the definition of $E^1$-page and its alternative interpretation in terms of the bulk-boundary correspondence, $\text{Ker}(d^{1}_{p,-q})$ can be understood as $q$-dimensional SPT phases on $p$-cells that create no anomaly on adjacent $(p-1)$-cells, and $\text{Im}(d^{1}_{p+1,-q})$ denotes the $q$-dimensional SPT phases on $p$-cells that can be constructed by adiabatically pumping the SPT states in adjacent $(p+1)$-cells. Therefore, $E^{2}_{p,-q}$ represents a set of $q$-dimensional SPT phases on $p$-cells that can extend to adjacent $(p-1)$-cells without creating anomalies and cannot be trivialized by the SPT states in adjacent $(p+1)$-cells. We can generalize this idea to define the higher differentials and the $E^r$-page
\bea\label{r-differential and Er+1-page}
\begin{aligned}
&d^{r}_{p,-q}:E^{r}_{p,-q}\rightarrow E^{r}_{p-r,-q+r-1},\\
&E^{r+1}_{p,-q}:=\text{Ker}(d^{r}_{p,-q})/\text{Im}(d^{r}_{p+r,-q-r+1}).
\end{aligned}
\eea
The $r$-th differential $d^{r}_{p,-q}$ satisfies the relation $d^{r}_{p,-q} \circ d^{r}_{p+r,-q-r+1} = 0$, which allows us to define the $E^{r+1}$-page. The physical meaning of $E^{r+1}_{p,-q}$ is that it represents a set of $q$-dimensional SPT phases on $p$-cells that can anomaly-freely extend to adjacent lower-dimensional cells (down to $(p-r)$-cells) and cannot be trivialized by the SPT states in adjacent higher-dimensional cells (up to $(p+r)$-cells). Thus, for a $d$-dimensional space manifold $X$, the $E^{r}$-page converges at the $E^{d+1}$-page, which corresponds to the limiting page $E^{\infty}$ that was introduced earlier. In conclusion, by using the AHSS, we can begin with the $E^1$-page and proceed step by step until reaching the limiting page $E^{\infty}$. Then, we can compute the generalized homology $h_{n}^{G}(X,Y)$ by solving a set of short exact sequence~\eqref{short exact sequences}.

\subsection{Extensive trivialization in the AHSS}
To illustrate the AHSS from a more physical perspective, let’s begin by considering a crystalline SPT state $\ket{\psi}$, which is the unique ground state of a $G$-symmetric system, where $G$ represents a crystalline point group symmetry. 
Suppose that without the constraint of the symmetry $G$, $\ket{\psi}$ can be adiabatically deformed into a trivial product state $\ket{T}$ through a finite-depth local unitary transformation $U^{loc}$: $U^{loc}\ket{\psi}=\ket{T}$. For such cases, if $\ket{\psi}$ is a non-trivial crystalline SPT state, no $U^{loc}$ can trivialize it while preserving the symmetry $G$. Nevertheless, it is still possible to define a local unitary transformation to extensively trivialize $\ket{\psi}$ while preserving $G$~\cite{constructionCSPTfromlowerd1}. By dividing the system into multiple regions under the $G$-symmetric constraint, we can apply a local unitary transformation that acts solely on a region $\sigma$, denoted $U^{loc}_{\sigma}$, to trivialize the state in $\sigma$. Specifically, the action of $U^{loc}_{\sigma}$ on $\ket{\psi}$ can be written as
\bea
U^{loc}_{\sigma}\ket{\psi}=\ket{T_{\sigma}}\otimes\ket{\psi_{\bar{\sigma}}},
\eea
where $\ket{T_{\sigma}}$ is the trivial product state in the region $\sigma$, and $\ket{\psi_{\bar{\sigma}}}$ represents the state in the remainder of the system, $\bar{\sigma}$. To extensively trivialize $\ket{\psi}$ in a symmetric way, we need to trivialize the states in regions $g\sigma$ by $U^{loc}_{g\sigma}=U_{g}U^{loc}_{\sigma}U^{-1}_{g}$ simultaneously, where $g\in G$ and $U_{g}$ is the unitary operator representing the action of $g$. Such a symmetric trivialization results in
\bea
\prod_{g\in G} U^{loc}_{g\sigma}\ket{\psi}=\bigotimes_{g\in G}\ket{T_{g\sigma}}\otimes\ket{\psi_{rest}},
\eea
where $\ket{\psi_{rest}}$ denotes the state in the remaining part of the system (the complement of $\cup_{g}g\sigma$). Consequently, the topological properties of $\ket{\psi}$ are encoded in $\ket{\psi_{rest}}$.

One systematic way to symmetrically divide a system into several regions is through the $G$-symmetric cell decomposition discussed in Sec.~\ref{Cell decomposition and the AHSS}. Therefore, the extensive trivialization can be implemented based on the cell decomposition, aligning naturally with the framework of the AHSS. To illustrate this connection, let’s first recall that the classification of crystalline SPT phases can be determined by
\bea\label{short exact sequence for ground state}
0\rightarrow F_{d-1}h_{0}\rightarrow h_{0}^{G}(X,Y)\rightarrow E^{\infty}_{d,-d}\rightarrow 0.
\eea
As mentioned before, $F_{d-1}h_{0}$ represents a set of $(d-1)$-dimensional SPT phases on the skeleton $X_{d-1}$, and $E^{\infty}_{d,-d}$ can be viewed as a set of $d$-dimensional SPT phases on $D^{d}=\cup_j D^{d}_j$. If the state $\ket{\psi}$ is trivial with respect to $E^{\infty}_{d,-d}$, we can symmetrically divide the system into two regions, $X_{d-1}$ and $D^{d}$, and then extensively trivialize it as
\bea
U^{loc,G}\ket{\psi}=\ket{\psi_{X_{d-1}}}\otimes\ket{T_{D^{d}}},
\eea
where $U^{loc,G}$ is a $G$-symmetric local unitary transformation, $\ket{\psi_{X_{d-1}}}$ is the state in the skeleton $X_{d-1}$, and $\ket{T_{D^{d}}}$ is the trivial product state in $D^{d}$. As a result, the crystalline SPT phases of the system can be fully represented by the SPT phases on $X_{d-1}$. 
When the state $\ket{\psi}$ is not trivial with respect to $E^\infty_{d,-d}$ but a torsion element of $E^\infty_{d,-d}$, 
i.e., $\ket{\psi}^{\otimes m}$ is trivial with respect to $E^\infty_{d,-d}$ for some integer $m$, then $\ket{\psi}^{\otimes m}$ is extensively trivialized as 
\bea
U^{loc,G}\ket{\psi}^{\otimes m}=\ket{\psi_{X_{d-1}}}\otimes\ket{T_{D^{d}}}. 
\eea
If the state $\ket{\psi_{X_{d-1}}}$ is nontrivial with respect to $F_{d-1}h_{0}$, the extension (\ref{short exact sequence for ground state}) is nontrivial. 
Moreover, when the state $\ket{\psi}$ is in an integer class of $E^{\infty}_{d,-d}$, $\ket{\psi}$ does not contribute to a nontrivial extension of (\ref{short exact sequence for ground state}).

\subsection{The AHSS for free-fermion systems}
For the free-fermion case, the generalized homology $h_{n}^{G}(X,Y)$ is reduced to the K-homology $K_n^{G}(X,Y)$, which gives the classification of free-fermion crystalline SPT phases. The AHSS can be employed to compute $K_n^{G}(X,Y)$ using a framework similar to that for the interacting case, but with some differences. In this context, the integer $n$ signifies a shift in the AZ symmetry class by adding $n$ chiral symmetries in addition to $G$. Moreover, $E^{r+1}_{p,-q}$ and $d^{r}_{p,-q}$ differ slightly from those in interacting systems. Rather than describing $q$-dimensional SPT phases on $p$-cells, they now pertain to SPT phases with the AZ class $(p+q)$ on $p$-cells, which we denote as $(p+q)$-th graded SPT phases.

Here, we briefly introduce how to classify crystalline SPT phases without additional internal symmetry using the AHSS. Due to Bott periodicity, $E^{r+1}_{p,-q}$ with odd/even $(p+q)$ represents a set of SPT phases within the class AIII/A on $p$-cells. In particular, the $E^1$-page can be determined by counting the topological number of states in each irreducible representation (irrep) of the little group $G_{D^{p}_j}$, leading to the following page
\bea
\centering
\renewcommand{\arraystretch}{1.5}
\begin{tabular}{c|ccccc}
$q=0$        &  Class A & Class AIII &  Class A & Class AIII & $\ldots$ \\
$q=1$        & $0$           & $0$ & $0$ & $0$ & $\ldots$ \\ \hline
$E^{1}_{p,-q}$ & $p=0$         & $p=1$ & $p=2$ & $p=3$ & $\ldots$ \\
\end{tabular}.
\eea
The non-zero elements of the above page take the form $\mathbb{Z}^m$, where $m$ is the number of irreps of $G_{D^{p}_j}$. One can begin with this $E^1$-page and proceed with the framework introduced in Sec.~\ref{Cell decomposition and the AHSS} to calculate $K_n^{G}(X,Y)$.

Notably, free-fermion crystalline SPT phases can be classified using twisted equivariant K-theory~\cite{TKtheory1,TKtheory2} as well, where the AHSS is likewise applicable~\cite{freeAHSSinkspace}. It is important to emphasize that mathematical duality asserts that K-theory and K-homology provide equivalent classifications.

\subsection{Relating free-fermion and interacting systems through the AHSS}
The $K$-group that classifies free fermion SPT phases is represented by the one-particle Hamiltonian ${\cal H}$. The state $\ket{\psi^{\rm free}} = \sum_{E_n < 0} \chi^\dag_n \ket{0}$, which is the Fermi sea of all the states with the negative energy of ${\cal H}$, is well-defined even in the many-body Fock space. 
Hence, a homomorphism $\kappa: K^G_0(X,Y) \to h^G_0(X,Y)$ can be defined. 
Physically, it is reasonable to assume that the homomorphism $\kappa$ is compatible with the boundary homomorphism, i.e., the bulk-boundary correspondence $\partial: h^G_0(X,Y) \to h^G_{-1}(Y)$ and $\partial: K^G_0(X,Y) \to K^G_{-1}(Y)$, which can be expressed as:
\begin{align}
\partial \circ \kappa = \kappa \circ \partial.
\end{align}
As a result, a homomorphism from free fermion SPT phases to many-body SPT phases can also be defined for each ingredient of the AHSS as well, and the differentials in AHSS commute with $\kappa$. 
Denoting the $E^r$-pages of the free fermion SPT phases and many-body SPT phases as $E^{{\rm free},r}_{p,-p}$ and $E^r_{p,-p}$, respectively, we have 
\begin{align}
\label{kappa_er}
\kappa^r_{p,-p}: E^{{\rm free},r}_{p,-p} \to E^r_{p,-p}.
\end{align}
Since the $E^r$-pages $E^r_{p,-p}$ and $E^{{\rm free},r}_{p,-p}$ are sublattices of the $E^1$-pages $E^1_{p,-p}$ and $E^{{\rm free},1}_{p,-p}$, the homomorphism \eqref{kappa_er} is determined by $\kappa^1_{p,-p}$ between the $E^1$-pages. Therefore, the explicit calculation of the homomorphism for $r=1$,
\begin{align}
\label{kappa_er_1}
\kappa^1_{p,-p}: E^{{\rm free},1}_{p,-p} \to E^1_{p,-p},
\end{align}
becomes the main focus.

In this paper, we restrict our discussion to cases where nontrivial SPT phases do not exist on 1- and 3-cells, i.e., $E^{{\rm free},1}_{p,-p} = E^1_{p,-p} = 0$ for $p=1,3$. Under this assumption, the short exact sequence \eqref{short exact sequences} and the homomorphism $\kappa$ can be summarized in the following commutative diagram:
\begin{align}
\label{kappa_ess}
\begin{CD}
0 @>>> E^{{\rm free},2}_{0,0} @>>> K^G_0 @>>> E^{{\rm free},2}_{2,-2} @>>> 0 \\
@. @V\kappa^2_{0,0}VV @V\kappa VV @V\kappa^2_{2,-2}VV \\
0 @>>> E^2_{0,0} @>>> h^G_0 @>>> E^2_{2,-2} @>>> 0 \\
\end{CD}
\end{align}
Thus, explicit calculations of $\kappa^1_{p,-p}$ for $p=0,2$ are of interest. 
For $p=0$, the zero-dimensional case, $\kappa^1_{0,0}$ can be easily computed. 
For $p=2$, we consider cases where there is no reduction, i.e., $\kappa^1_{2,-2}$ is an isomorphism. 
Therefore, once the horizontal extensions of the short exact sequence \eqref{kappa_ess} are determined, the reduction $\kappa: K^G \to h^G_0$ is fully specified.

The notation used in this paper is as follows. 
For the $E^1$ page in the form
\begin{align}
E^1_{p,-p} = \bigoplus_{j=1}^q \mathbb{Z}_{k_j},
\end{align}
where $\mathbb{Z}_{k_j=\infty}$ means $\mathbb{Z}_\infty = \mathbb{Z}$, 
an element $(n_1, \dots, n_q) \in E^1_{p,-p}$ correspond to tensor product states on $p$-cells as 
\begin{align}
\ket{\psi(n_1, \dots, n_q)} = \bigotimes_{i=1}^q \ket{\psi_j}^{\otimes n_j},
\end{align}
where $\ket{\psi_j}$ is the state on $p$-cells corresponding to the generator of the direct summand $\mathbb{Z}_{k_j}$. 
The same applies to the $E^1$-page of free fermion SPT phases, $E^{{\rm free},1}_{p,-p}$. 
To distinguish free fermion and many-body SPT phases, subscripts are used, such as $(n_1, \dots, n_q)_f \in E^{{\rm free},1}_{p,-p}$ and $(n_1, \dots, n_q)_I \in E^1_{p,-p}$.
Note that $(n_1, \dots, n_q)$ does not represent topological invariants, such as the filling number or Chern number defined from wave functions. 
Instead, it denotes the exponents in the decomposition of states defined only on the $p$-cell into the tensor product of generators of the $E^1$-page.
When it is clear from the context, the superscript ``free'' is omitted.

\section{\texorpdfstring{1$d$}{\textmu} systems with \texorpdfstring{$U(1)$}{\textmu} and reflection symmetry}\label{Main reflection Sec}
\begin{figure}[htp!]
\centering
\subfloat[]{\includegraphics[width=0.35\textwidth]{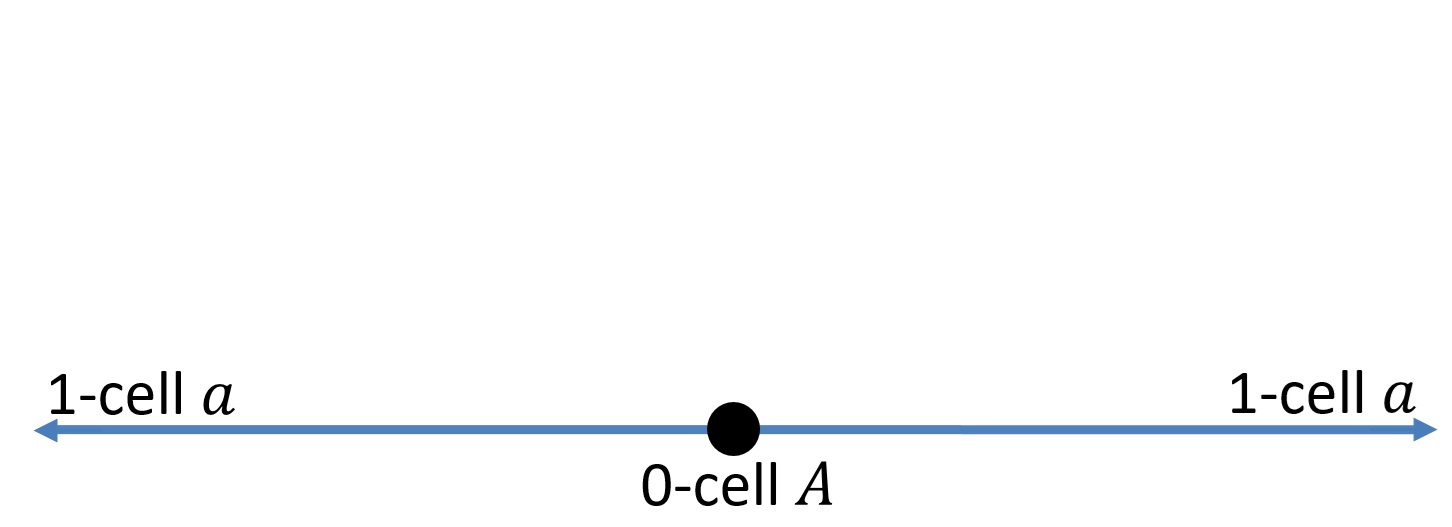}\label{Cell_decomposition_reflection}}\hskip 0.5cm
\subfloat[]{\includegraphics[width=0.35\textwidth]{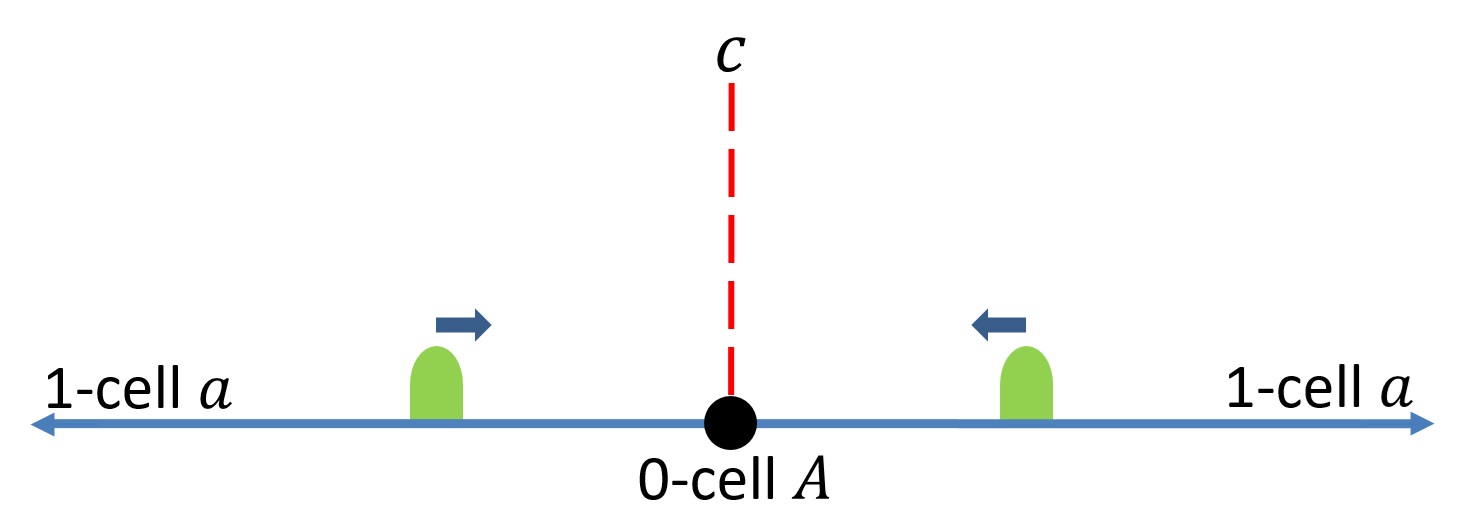}\label{First_differential_reflection}}\\
\caption{(a) The $\mathbb{Z}_2$-symmetric cell decomposition of $1d$ space. The 0-cell $A$ has $U(1)$ and reflection symmetry, while the 1-cell $a$ has $U(1)$ symmetry only. (b) The first differential $d^{1}_{1,0}$ can be regarded as pumping two SPT states in 1-cell $a$ onto the reflection center (0-cell $A$) while preserving reflection symmetry. The red dashed line denotes the reflection center, and the green wave packets represent the SPT states in the 1-cell $a$.}
\end{figure}
For $1d$ systems with $U(1)$ and reflection symmetry, the free-fermion topological classification can be given by the $K$-homology $K_{0}^{\mathbb{Z}_2}(\mathbb{R},\partial \mathbb{R})$. In the interacting scope, the generalized homology $h_{0}^{\mathbb{Z}_2}(\mathbb{R},\partial \mathbb{R})$ provides the classification of SPT phases. Utilizing eq.~\eqref{short exact sequences}, one can fit the homology $K_{0}^{\mathbb{Z}_2}(\mathbb{R},\partial \mathbb{R})$ and $h_{0}^{\mathbb{Z}_2}(\mathbb{R},\partial \mathbb{R})$ into the following short exact sequence:
\bea\label{short exact sequence for 1d reflection}
0\rightarrow E^{2}_{0,0}\rightarrow \text{$K_{0}^{\mathbb{Z}_2}(\mathbb{R},\partial \mathbb{R})$ or $ h_{0}^{\mathbb{Z}_2}(\mathbb{R},\partial \mathbb{R})$}\rightarrow E^{2}_{1,-1}\rightarrow 0.
\eea
In the latter discussion, we will show that $E^{2}_{1,-1}=0$ and hence 
\bea
\text{$K_{0}^{\mathbb{Z}_2}(\mathbb{R},\partial \mathbb{R})$ or $ h_{0}^{\mathbb{Z}_2}(\mathbb{R},\partial \mathbb{R})$}\cong E^{2}_{0,0}\cong  E^{1}_{0,0}/\text{Im}(d^{1}_{1,0}).
\eea
As mentioned in Sec.~\ref{Cell decomposition and the AHSS}, the first differential $d^{1}_{1,0}:E^{1}_{1,0}\rightarrow E^{1}_{0,0}$ can be interpreted as pumping the SPT states from the 1-cell onto the 0-cell, as shown in Fig.~\ref{First_differential_reflection}. Consequently, the quotient $E^{1}_{0,0}/\text{Im}(d^{1}_{1,0})$ represents a set of SPT phases on the 0-cell that cannot be trivialized by SPT states in the 1-cell. In the remainder of this section, we will demonstrate that both the free-fermion and interacting topological classifications can be described in terms of the quantum numbers associated with the SPT phases on the $0$-cell. By examining the relation between the quantum numbers in the free-fermion and interacting settings, we can then establish the connection between the topological equivalence classes in these two contexts.
\subsection{Free-fermion crystalline SPT phases}\label{1d R free SPT sec}
Let's begin with the free-fermion SPT phases for 1$d$ systems with $U(1)$ and reflection symmetry. After performing the $\mathbb{Z}_2$-symmetric cell decomposition as shown in Fig.~\ref{Cell_decomposition_reflection}, we obtain the following $E^1$-page~\cite{freeAHSSinrealspace}
\bea\label{E-1 page, 1d free reflection}
\centering
\renewcommand{\arraystretch}{1.5}
\begin{tabular}{c|cc}
$q=0$        &  $\mathbb{Z}^2$ & $\mathbb{Z}$ \\
$q=1$        & $0$           & $0$ \\ \hline
$E^{1}_{p,-q}$ & $p=0$         & $p=1$   \\
\end{tabular}.
\eea
$E^{1}_{0,0}$ describes the number of states in the irreps with reflection eigenvalues $+1$ and $-1$, respectively. $E^{1}_{1,0}$ counts the number of gapless edge states of a 1$d$ topological insulator within Class AIII in the 1-cell $a$. Therefore, the first differential $d^{1}_{1,0}$ can be interpreted as the process of pumping the 1st-graded SPT states in the 1-cell $a$ (the topological gapless edge states of the 1$d$ chiral-symmetric topological insulator in the 1-cell $a$) onto the 0-cell $A$ while preserving reflection symmetry. In essence, $d^{1}_{1,0}$ maps the representations of the little group $G_{D^{1}_j}$ at the 1-cell, to the representation of the little group $G_{D^{0}_j}$ at the 0-cell, where $G_{D^{1}_j}\subset G_{D^{0}_j}$. This is nothing but the construction of the induced representation \cite{interactingAHSS}. To get a better understanding of this pumping process, we introduce the quantum number $(n_{+}, n_{-})_f$ to describe the $\mathbb{Z}^2$ classification on the 0-cell $A$, where $n_{+}$ and $n_{-}$ denote the number of states (in the 0-cell) with reflection eigenvalues $+1$ and $-1$, respectively, and the subscript $f$ indicates the consideration of free-fermion systems. With this quantum number, the first differential (induced representation) can be written as
\bea
d^{1}_{1,0}: 1\rightarrow (1,1)_f,
\eea
which leads to
\bea
E^{2}_{0,0}\cong E^{1}_{0,0}/\text{Im}(d^{1}_{1,0})= \mathbb{Z}^2/\mathbb{Z}(1,1)_f=\mathbb{Z}.
\eea
By definitions~\eqref{1-differential} and~\eqref{E2-page}, it is clear that $E^{2}_{1,-1}=0$ because $E^{1}_{1,-1}=0$. Substituting these results into eq.~\eqref{short exact sequence for 1d reflection}, we obtain
\bea
K_{0}^{\mathbb{Z}_2}(\mathbb{R},\partial \mathbb{R})\cong E^{2}_{0,0}\cong E^{1}_{0,0}/\text{Im}(d^{1}_{1,0})=\mathbb{Z}.
\eea
In fact, $K_{0}^{\mathbb{Z}_2}(\mathbb{R},\partial \mathbb{R})\cong E^{1}_{0,0}/\text{Im}(d^{1}_{1,0})$ in the above equation illustrates a bulk-boundary correspondence. This correspondence indicates that the topological classification of the 1$d$ systems can be determined by the SPT phases on the reflection center (the 0-cell $A$), which can be described by the quantum number $(n_{+},n_{-})_f$ trivialized by the following equivalence relation
\bea
(n_{+},n_{-})_f+(1,1)_f \sim (n_{+},n_{-})_f.
\eea
By using the above relation, we have
\bea
(n_{+},n_{-})_f\sim (n_{+},n_{-})_f-n_{-}(1,1)_f = (n_{+}-n_{-},0)_f.
\eea
In terms of the quantum number, the $\mathbb{Z}$ classification of these free-fermion SPT phases can be generated by either $(1,0)_f$ or $(0,1)_f$. If we select $(1,0)_f$ as the generator, the above equation indicates that the equivalence classes of this $\mathbb{Z}$ classification can be expressed as
\bea\label{1d Z class and n}
[n_{+}-n_{-}].
\eea
\subsection{Interacting crystalline SPT phases}\label{1d R interacting SPT sec}
For 1$d$ interacting systems with $U(1)$ and reflection symmetry, the relevant $E^1$-page is given by\cite{interactingAHSS}
\bea\label{E-1 page, 1d interacting reflection}
\centering
\renewcommand{\arraystretch}{1.5}
\begin{tabular}{c|cc}
$q=0$        &  $\mathbb{Z}\times \mathbb{Z}_2$ & $\mathbb{Z}$ \\
$q=1$        & $0$                              & $0$ \\ \hline
$E^{1}_{p,-q}$ & $p=0$                          & $p=1$   \\
\end{tabular}.
\eea
Here, $E^{1}_{0,0}$ characterizes the integer-valued $U(1)$ charge and the reflection eigenvalue of a SPT state $\ket{\psi}=\psi^{\dag}\ket{0}$ on the 0-cell, where $\ket{0}$ represent the Fock vacuum. Hence, the $\mathbb{Z}\times \mathbb{Z}_2$ classification can be described by the quantum number $(N_C,R_C)_I$, where
\bea\label{NCRC reflection def}
N_C=\bra{\psi}\hat{N}\ket{\psi}\in \mathbb{Z}, \quad R_C=\left\{ 
\begin{aligned}
0,\quad&\text{if} \ R\psi^{\dag} R^{-1}=\psi,&\\
1,\quad&\text{if} \ R\psi^{\dag} R^{-1}=-\psi,&
\end{aligned}
\in \mathbb{Z}_2.
\right.
\eea
$R$ is the reflection operator, and the subscript $I$ emphasizes the consideration of interacting systems. The first differential $d^{1}_{1,0}$ describes how SPT phases on the reflection center are constructed by pumping $0d$ SPT states from the 1-cell $a$ onto the reflection center. Specifically, this process can be viewed as pumping two complex fermions, $f^{\dag}_1$ and $f^{\dag}_2$, with the property $Rf^{\dag}_{1}R^{-1}=f^{\dag}_2$, onto the reflection center. Since we have $Rf^{\dag}_{1}f^{\dag}_{2}R^{-1}=f^{\dag}_{2}f^{\dag}_{1}=-f^{\dag}_{1}f^{\dag}_{2}$, the first differential in terms of $(N_C,R_C)_I$ is
\bea
d^{1}_{1,0}: 1\rightarrow (2,1)_I,
\eea
so we obtain
\bea
E^{2}_{0,0}\cong E^{1}_{0,0}/\text{Im}(d^{1}_{1,0})=\mathbb{Z}\times \mathbb{Z}_2/\mathbb{Z}(2,1)_I=\mathbb{Z}_4.
\eea
For the same reason as in the free-fermion case, $E^{2}_{1,-1}=0$ due to $E^{1}_{1,-1}=0$. Taking all these into account, the classification is given by
\bea
h_{0}^{\mathbb{Z}_2}(\mathbb{R},\partial \mathbb{R})\cong E^{2}_{0,0}\cong E^{1}_{0,0}/\text{Im}(d^{1}_{1,0})=\mathbb{Z}_4.
\eea

Similar to the concept discussed for the free-fermion case, the topological classification here can be characterized by the quantum number $(N_C, R_C)_I$ trivialized by the following equivalence relation:
\bea
(N_C,R_C)_I+(2,1)_I\sim (N_C,R_C)_I,
\eea
which leads to
\bea
(N_C,R_C)_I\sim (N_C,R_C)_I-R_C(2,1)_I= (N_C-2R_C,0)_I\sim(N_C-2R_C\,\text{mod}\, 4,0)_I.
\eea
Given that the $\mathbb{Z}_4$ classification can be generated by $(1,0)_I$, a system with the quantum number $(N_C,R_C)_I$ falls into the class
\bea
[N_C-2R_C\,\text{mod}\, 4].
\eea

\subsection{Interaction effects on free-fermion crystalline SPT phases}\label{free to interaction, reflection}
\begin{table}[htb!]
\centering
\renewcommand{\arraystretch}{1.5}
\begin{tabular}{c|c|c}
  Symmetry group  & Free-fermion SPT phases & Interacting SPT phases  \\ \hline 
$U(1)\times \mathbb{Z}_2^{R}$ & $[N_R]$ & $[N_R \,\text{mod}\,4]$ \\ 
\end{tabular}
\caption{The relation between free-fermion and interacting SPT phases for $1d$ systems with $U(1)$ and reflection symmetry. Here, $N_R=n_+-n_-$.}
\end{table}
Without loss of generality, a free-fermion SPT state on the 0-cell can be considered as a $0d$ single-particle state, which can be viewed as a single complex fermion (electron). Adopting this concept, one can see how the interaction effects reduce the free-fermion SPT phases. We start by labeling a $0d$ single-particle state, or equivalently a fermion, with reflection eigenvalues $+1$ and $-1$ as $\phi^{\dag}_{+}$ and $\phi^{\dag}_{-}$, respectively. As mentioned previously, the free-fermion SPT phases can be described by the quantum number $(n_{+},n_{-})_f$, which corresponds to the class $[n_{+}-n_{-}]$. In this context, $n_{+}$ and $n_{-}$ is the number of $\phi^{\dag}_{+}$ and $\phi^{\dag}_{-}$. Thus, a system in the class $[n_{+} - n_{-}]$ can be represented by a multi-fermionic system consisting of $n_{+}$ fermion $\phi^{\dag}_{+}$ and $n_{-}$ fermion $\phi^{\dag}_{-}$. This explanation helps us clarify the relation between free-fermion and interacting classification. Since the fermionic Fock state of a multi-fermionic system is essentially a Slater determinant, a free-fermion system within the class $[n_{+}-n_{-}]$ in the Fock space can be represented by the following SPT state (in the 0-cell)
\bea
\ket{\psi}=\prod^{n_{+}}_{j=1} \phi^{\dag}_{+,j} \prod^{n_{-}}_{l=1} \phi^{\dag}_{-,l} \ket{0}.
\eea
Now, we can assign the quantum number $(N_C,R_C)_I$ to free-fermion systems. By the definition \eqref{NCRC reflection def}, and given that $R\phi^{\dag}_{+,j}R^{-1}=\phi^{\dag}_{+,j}$ and $R\phi^{\dag}_{-,l}R^{-1}=-\phi^{\dag}_{-,l}$,the corresponding quantum number $(N_C,R_C)_I$ for a free-fermion system within the class $[n_{+}-n_{-}]$ is
\bea
(n_{+}+n_{-},n_{-}\,\text{mod}\,2)_I.
\eea
With the equivalence relation $(N_C,R_C)_I + (2, 1)_I \sim (N_C, R_C)_I$, the quantum number can be further simplified to
\bea
(n_{+}+n_{-},n_{-}\,\text{mod}\,2)_I-n_{-}(2,1)_I=(n_{+}-n_{-},0)_I \sim (n_{+}-n_{-}\,\text{mod}\,4,0)_I.
\eea
As stated in Sec.~\ref{1d R interacting SPT sec}, since the $\mathbb{Z}_4$ classification is generated by $(1,0)_I$, the interacting class into which this free-fermion system falls is $[n_{+}-n_{-}\,\text{mod}\,4]$. In summary, if a $1d$ free-fermion system with $U(1)$ and reflection symmetry falls into a class $[N_R]$, the interaction effect will reduce the class to $[N_R,\text{mod},4]$,
\bea\label{reflection interaction effect}
[N_R]\xrightarrow{\text{Interaction effect}}[N_R\,\text{mod}\,4],
\eea
where $N_R=n_{+}-n_{-}$. In Appendix~\ref{Hpn and SPT phases}, we present a series of lattice models that clearly demonstrate the connection between the free-fermion and interacting classes we establish here.

\section{\texorpdfstring{2$d$}{\textmu} systems with \texorpdfstring{$U(1)$}{\textmu} and \texorpdfstring{$C_n$}{\textmu} rotation symmetry}\label{Main Cn rotation sec}
\begin{figure}[htp!]
\centering
\subfloat[]{\includegraphics[width=0.35\textwidth]{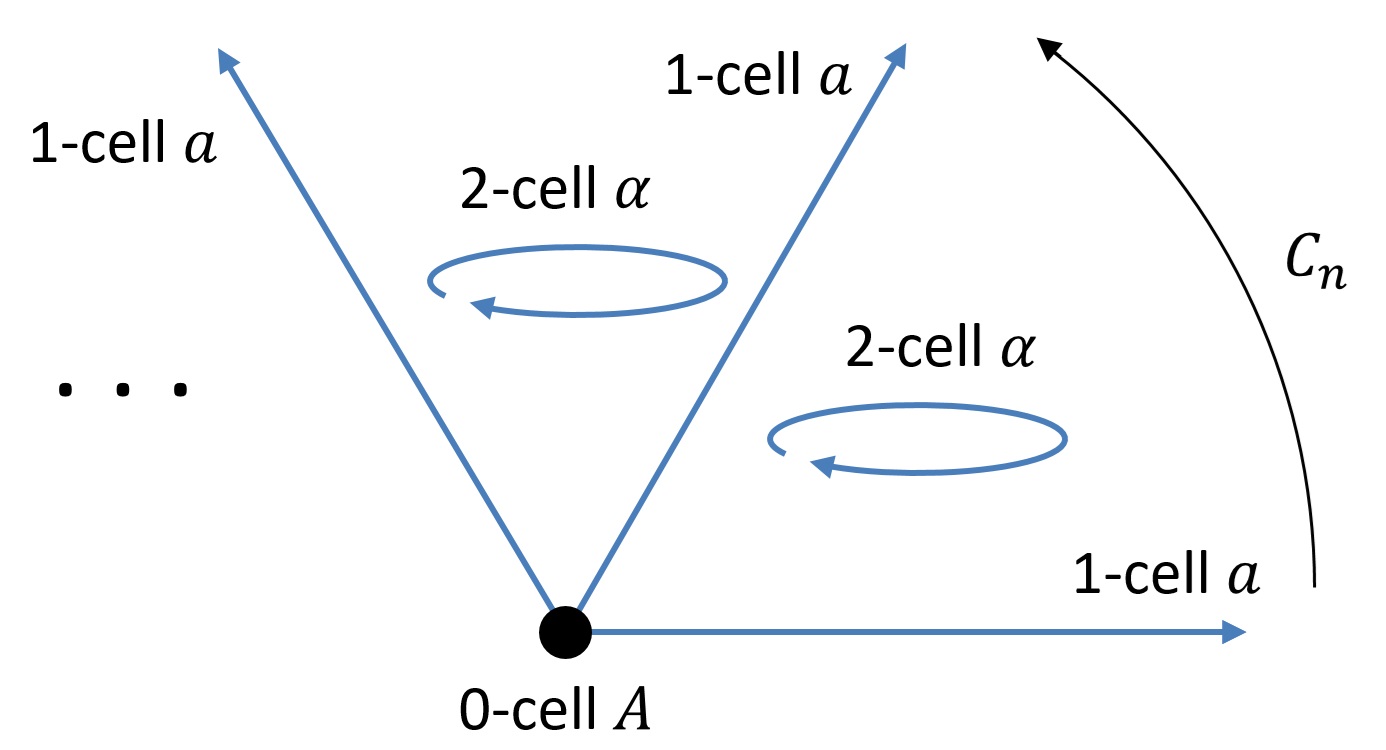}\label{Cell_decomposition_rotation}}\hskip 0.5cm
\subfloat[]{\includegraphics[width=0.35\textwidth]{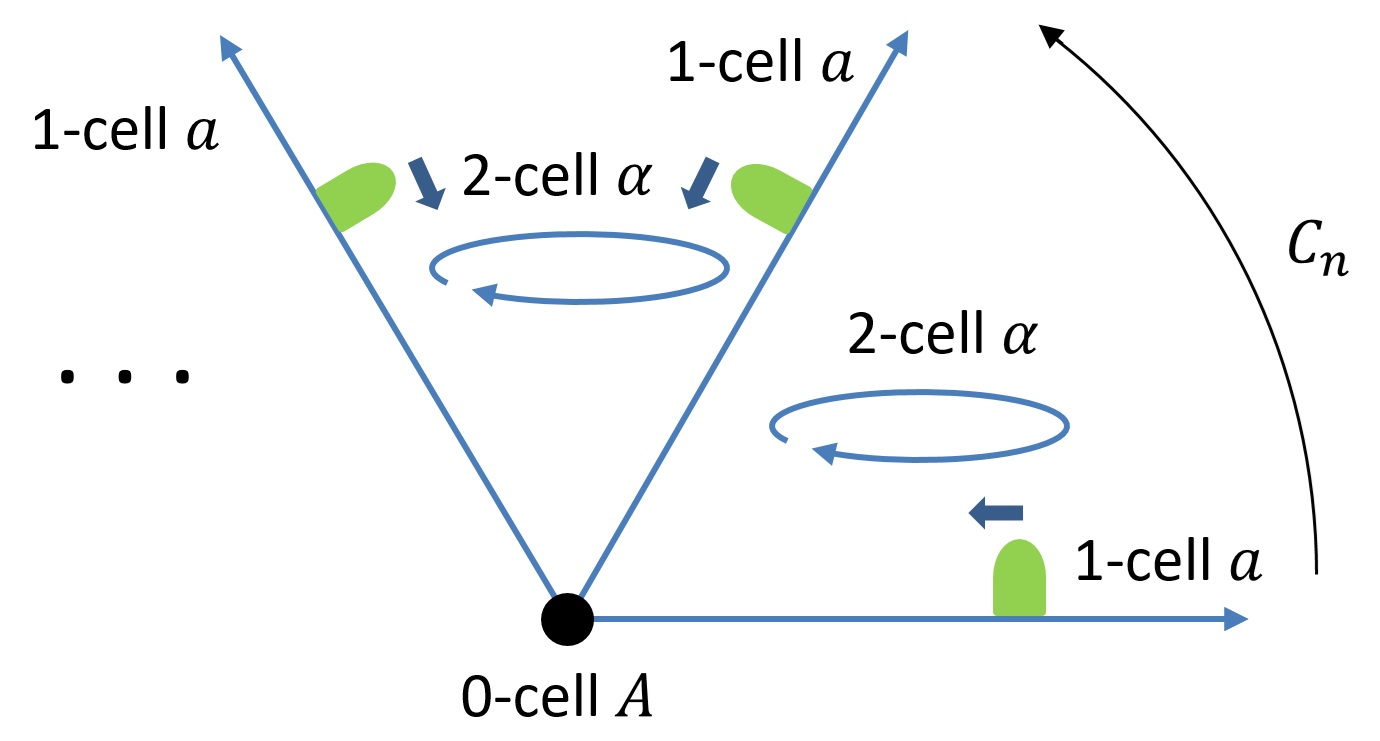}\label{First_differential_rotation}}\\
\caption{(a) The $\mathbb{Z}_n$-symmetric cell decomposition of $2d$ space. The 0-cell $A$ has $U(1)$ and $C_n$ rotation symmetry, while the 1-cell $a$ and the 2-cell $\alpha$ have $U(1)$ symmetry only. (b) The first differential $d^{1}_{1,0}$ under the $C_n$ rotation symmetry.}
\end{figure}

Let's consider the $2d$ systems with $U(1)$ and $C_n$ rotation symmetry. Their free-fermion and interacting classification can be determined by the $K$-homology $K_{0}^{\mathbb{Z}_n}(\mathbb{R}^2,\partial \mathbb{R}^2)$ and the generalized homology $h_{0}^{\mathbb{Z}_n}(\mathbb{R}^2,\partial \mathbb{R}^2)$, respectively. In the framework of the AHSS, $K_{0}^{\mathbb{Z}_n}(\mathbb{R}^2,\partial \mathbb{R}^2)$ and $h_{0}^{\mathbb{Z}_n}(\mathbb{R}^2,\partial \mathbb{R}^2)$ can be fitted into the following short exact sequences
\bea\label{short exact sequence for 2d rotation}
\begin{alignedat}{3}
0\rightarrow &  E^{3}_{0,0}&&\rightarrow \quad\quad\quad\quad\,F_{1}h_{0}&&\rightarrow E^{3}_{1,-1}\rightarrow 0,\\
0\rightarrow &  F_{1}h_{0}&&\rightarrow  \text{$K_{0}^{\mathbb{Z}_n}(\mathbb{R}^2,\partial \mathbb{R}^2)$ or $ h_{0}^{\mathbb{Z}_n}(\mathbb{R}^2,\partial \mathbb{R}^2)$}&&\rightarrow E^{3}_{2,-2}\rightarrow 0.\\
\end{alignedat}
\eea
We will demonstrate that the solution to the group extension problems is
\bea
\text{$K_{0}^{\mathbb{Z}_n}(\mathbb{R}^2,\partial \mathbb{R}^2)$ or $ h_{0}^{\mathbb{Z}_n}(\mathbb{R}^2,\partial \mathbb{R}^2)$}\cong E^{2}_{0,0}\oplus E^{2}_{2,-2}\cong E^{1}_{0,0}/\text{Im}(d^{1}_{1,0})\oplus E^{1}_{2,-2}
\eea
Similar to the concept mentioned in Sec.~\ref{Main reflection Sec}, $E^{1}_{0,0}$ describes the $0d$ SPT phases on the $0$-cell $A$. The first differential $d^{1}_{1,0}$ indicates how $0d$ SPT states in the 1-cell $a$ are pumped onto the reflection center (0-cell $A$), as shown in Fig.~\ref{First_differential_rotation}. $E^{1}_{2,-2}$ can be interpreted as representing Chern insulators (and bosonic integer quantum Hall states~\cite{BIQHE}) in the 2-cell $\alpha$. Notably, one can regard $E^{2}_{2,-2}\cong E^{1}_{2,-2}$ as a consequence of the fact that $C_n$ rotation does not change the Hall conductivity $\sigma_{xy}$ or the thermal Hall conductivity $\kappa_{xy}$.
\subsection{Free-fermion crystalline SPT phases}\label{SPT free roation}
For $2d$ free-fermion systems with $U(1)$ and $C_n$ rotation symmetry, the $\mathbb{Z}_n$-symmetric cell decomposition (Fig.~\ref{Cell_decomposition_rotation}) yields the following $E^1$-page \cite{freeAHSSinrealspace}
\bea\label{E-1 page, 2d free rotation}
\centering
\renewcommand{\arraystretch}{1.5}
\begin{tabular}{c|ccc}
$q=0$        &  $\mathbb{Z}^n$ & $\mathbb{Z}$ & $\mathbb{Z}$\\
$q=1$        & $0$           & $0$ & $0$\\ \hline
$E^{1}_{p,-q}$ & $p=0$         & $p=1$ & $p=2$   \\
\end{tabular}.
\eea 
By applying eq.~\eqref{1-differential} and eq.~\eqref{E2-page}, we get the $E^2$-page
\bea\label{E-2 page, 2d free rotation}
\centering
\renewcommand{\arraystretch}{1.5}
\begin{tabular}{c|ccc}
$q=0$        & $\mathbb{Z}^{n-1}$ & $0$ & $\mathbb{Z}$\\
$q=1$        & $0$           & $0$ & $0$\\ \hline
$E^{2}_{p,-q}$ & $p=0$         & $p=1$ & $p=2$   \\
\end{tabular}.
\eea
$E^{2}_{0,0}\cong E^{1}_{0,0}/\text{Im}(d^{1}_{1,0})$can be understood as follows: since $E^{1}_{0,0}$ pertains to the number of states in the irrep with rotation eigenvalue $e^{\gamma \frac{2\pi i}{n}}$, where $\gamma=0,1,\ldots,n-1$, it can be characterized by the quantum number
\bea
(n_0,n_1,\ldots,n_{n-1})_f,
\eea
where $n_{\gamma}$ denotes the number of SPT states (in the 0-cell) with the rotation eigenvalue $e^{\gamma \frac{2\pi i}{n}}$. Using this quantum number, the first differential—understood as pumping SPT states from the 1-cell onto the 0-cell while preserving rotation symmetry, as illustrated in Fig.~\ref{First_differential_rotation}—can be written as
\bea
d^{1}_{1,0}: 1\rightarrow (1,1,\ldots,1)_f,
\eea
and we have
\bea
E^{2}_{0,0}\cong E^{1}_{0,0}/\text{Im}(d^{1}_{1,0})=\mathbb{Z}^n/\mathbb{Z}(1,1,..,1)_f=\mathbb{Z}^{n-1}.
\eea
Since $C_n$ rotation does not affect the Hall conductivity or the thermal Hall conductivity, the differential $d^{1}_{2,0}=0$. Given this fact and the absence of $d^{1}_{3,-q}$, we have $E^{2}_{2,-2}=E^{2}_{2,0}\cong E^{1}_{2,0}=\mathbb{Z}$, where the $\mathbb{Z}$ classification represents the Chern number of Chern insulators in 2-cells. Lastly, due to the absence of higher differentials, the $E^3$-page is the same as $E^2$-page. Therefore, the solution to the short exact sequences~\eqref{short exact sequence for 2d rotation} is
\bea
\begin{aligned}
&F_{1}h_{0}=E^{2}_{0,0}\cong E^{1}_{0,0}/\text{Im}(d^{1}_{1,0})=\mathbb{Z}^{n-1}\\
&K_{0}^{\mathbb{Z}_n}(\mathbb{R}^2,\partial \mathbb{R}^2)\cong F_{1}h_{0}\oplus E^{1}_{2,-2}\cong E^{1}_{0,0}/\text{Im}(d^{1}_{1,0})\oplus E^{1}_{2,-2}=(\mathbb{Z}^{n-1})\times\mathbb{Z}.
\end{aligned}
\eea

Now, let’s use the same approach discussed in Sec.~\ref{Main reflection Sec} to assign quantum numbers to the classification. The $\mathbb{Z}^{n-1}$ classification is related to $E^{1}_{0,0}/\text{Im}(d^{1}_{1,0})$, implying a bulk-boundary correspondence: the SPT phases on the skeleton $X_1$ can be determined by the SPT phases on the 0-cell $A$. In terms of the quantum number $(n_0,n_1,\ldots,n_{n-1})_f$, this correspondence can be described by the following equivalence relation
\bea
(n_0,n_1,\ldots,n_{n-1})_f+(1,1,\ldots,1)_f\sim(n_0,n_1,\ldots,n_{n-1})_f.
\eea
Given that the $\mathbb{Z}^{n-1}$ classification can be generated by the set of quantum numbers $\{(0,1,0\ldots,0)_f,\break(0,0,1,0\ldots,0)_f,\ldots,(0,0,\ldots,0,1)_f\}$, using the equivalence relation, we have
\bea
(n_0,n_1,\ldots,n_{n-1})_f\sim(n_0,n_1,\ldots,n_{n-1})_f-n_0(1,1,\ldots,1)_f=(0,n_1-n_0,\ldots,n_{n-1}-n_0)_f,
\eea
which leads to the following equivalence classes
\bea
[(N_1,N_2,\ldots,N_{n-1})],
\eea
where $N_\gamma=n_\gamma-n_0$. For consistency, we use the quantum number $(N_{CI})_f$ to describe the $\mathbb{Z}$ classification for $E^{1}_{2,-2}$. Here, $N_{CI}$ represents the Chern number of Chern insulators in 2-cells, and a system with $(N_{CI})_f$ falls into class $[N_{CI}]$. Considering all these factors, the class of a free-fermion system with the quantum numbers $(n_0,n_1,\ldots,n_{n-1})_f$ and $(N_{CI})_f$ is
\bea
[(N_1,N_2,\ldots,N_{n-1})]\oplus[N_{CI}].
\eea

\subsection{Interacting crystalline SPT phases}
For $2d$ interacting systems with $U(1)$ and $C_n$ rotation symmetry, the corresponding $E^1$-page is \cite{interactingAHSS}
\bea\label{E-1 page, 2d interacting rotation}
\centering
\renewcommand{\arraystretch}{1.5}
\begin{tabular}{c|ccc}
$q=0$        &  $\mathbb{Z}\times \mathbb{Z}_n$ & $\mathbb{Z}$ & \\
$q=1$        & $0$           & $0$ & $0$\\ 
$q=2$        &  & $\mathbb{Z}^2$ & $\mathbb{Z}^2$\\ \hline
$E^{1}_{p,-q}$ & $p=0$         & $p=1$ & $p=2$   \\
\end{tabular}.
\eea
Certain elements in the above table are left blank, as they do not play a role in the construction of SPT phases. Since $E^{1}_{0,0}$ represents the integer-valued $U(1)$ charge and the rotation eigenvalue of a SPT state $\ket{\psi} = \psi^{\dag}\ket{0}$, one can use the quantum number $(N_C,R_C)_I$ to describe the $\mathbb{Z} \times \mathbb{Z}_n$ classification, where
\bea\label{NCRC rotation def}
N_C=\bra{\psi}\hat{N}\ket{\psi}\in \mathbb{Z}, \quad R_C=\gamma \in \mathbb{Z}_n\quad\text{for} \ C_n\psi^{\dag} C_n^{-1}=e^{\gamma \frac{2\pi i}{n}}\psi.
\eea
$C_n$ is the rotation operator. The first differential $d^{1}_{1,0}$ describes the construction that pumps the complex fermions $f^{\dag}_1$,$\ldots$, $f^{\dag}_n$, where $ C_n f^{\dag}_j C_n^{-1}=f^{\dag}_{j+1}$ and $f^{\dag}_{n+1}=f^{\dag}_{1}$, from the 1-cell $a$ onto the rotation center. For $n\in\text{odd}$, we have $C_n \Psi C_n^{-1}=\Psi$, where $\Psi=f^{\dag}_{1} f^{\dag}_{2}\ldots f^{\dag}_{n}$; for $n\in\text{even}$, the rotation operator acts as $C_n \Psi C_n^{-1}=-\Psi$. Hence, $d^{1}_{1,0}$ in terms of the quantum number can be written as
\bea
d^{1}_{1,0}: 1\rightarrow \left\{ 
\begin{aligned}
&(n,0)_I&,\quad&\text{for odd $n$,} \\
&(n,n/2)_I&,\quad&\text{for even $n$.}
\end{aligned}
\right.
\eea
On the other hand, the $\mathbb{Z}^2$ classification of $E^{1}_{1,-2}$ and $E^{1}_{2,-2}$ corresponds to Chern insulators ($\sigma_{xy}=1$, $\kappa_{xy}=1$) and Bosonic integer quantum Hall states  ($\sigma_{xy}=8$, $\kappa_{xy}=0$). Since the Hall conductivity and the thermal Hall conductivity remain unchanged under the action of $C_n$ rotation operators, we have $d^{1}_{2,-2}=0$. Using these results to compute the $E^2$-page, we obtain
\bea\label{E-2 page, 2d interacting rotation n=2}
\centering
\renewcommand{\arraystretch}{1.5}
\begin{tabular}{c|ccc}
$q=0$        &  $\mathbb{Z}_4$ & $0$ & \\
$q=1$        & $0$           & $0$ & $0$\\ 
$q=2$        &  &  & $\mathbb{Z}^2$\\ \hline
$E^{1}_{p,-q}$ & $p=0$         & $p=1$ & $p=2$   \\
\end{tabular},\quad\text{for $n=2$},
\eea
\bea\label{E-2 page, 2d interacting rotation odd n}
\centering
\renewcommand{\arraystretch}{1.5}
\begin{tabular}{c|ccc}
$q=0$        &  $\mathbb{Z}_n\times \mathbb{Z}_n$ & $0$ & \\
$q=1$        & $0$           & $0$ & $0$\\ 
$q=2$        &  &  & $\mathbb{Z}^2$\\ \hline
$E^{1}_{p,-q}$ & $p=0$         & $p=1$ & $p=2$   \\
\end{tabular},\quad\text{for odd $n$},
\eea
\bea\label{E-2 page, 2d interacting rotation even n}
\centering
\renewcommand{\arraystretch}{1.5}
\begin{tabular}{c|ccc}
$q=0$        &  $\mathbb{Z}_{2n}\times \mathbb{Z}_{n/2}$ & $0$ & \\
$q=1$        & $0$           & $0$ & $0$\\ 
$q=2$        &  &  & $\mathbb{Z}^2$\\ \hline
$E^{1}_{p,-q}$ & $p=0$         & $p=1$ & $p=2$   \\
\end{tabular},\quad\text{for even $n$}.
\eea
Since there are no higher differentials, the $E^3$-page is the same as $E^2$-page. Consequently, we have $F_{1}h_{0}=E^{2}_{0,0}\cong E^{1}_{0,0}/\text{Im}(d^{1}_{1,0})$ and
\bea
\begin{aligned}
h_{0}^{\mathbb{Z}_n}(\mathbb{R}^2,\partial \mathbb{R}^2)&\cong E^{1}_{0,0}/\text{Im}(d^{1}_{1,0})\oplus E^{1}_{2,-2}\\
&=\left\{ 
\begin{aligned}
&(\mathbb{Z}\times \mathbb{Z}_2)/\mathbb{Z}(2,1)_I\times\mathbb{Z}^2=(\mathbb{Z}_4)\times\mathbb{Z}^2&,\quad&\text{for $n=2$,} \\
&(\mathbb{Z}\times \mathbb{Z}_n)/\mathbb{Z}(n,0)_I\times\mathbb{Z}^2=(\mathbb{Z}_n\times\mathbb{Z}_n)\times\mathbb{Z}^2&,\quad&\text{for odd $n$,} \\
&(\mathbb{Z}\times \mathbb{Z}_n)/\mathbb{Z}(n,n/2)_I\times\mathbb{Z}^2=(\mathbb{Z}_{2n}\times\mathbb{Z}_{n/2})\times\mathbb{Z}^2&,\quad&\text{for even $n$.}
\end{aligned}
\right.
\end{aligned}
\eea

Here, in terms of the quantum number $(N_C,R_C)_I$, the bulk-boundary correspondence relevant to $E^{1}_{0,0}/\text{Im}(d^{1}_{1,0})$ is described by the following equivalence relation
\bea\label{equivalence relation for interacting rotation}
(N_C,R_C)_I\sim\left\{ 
\begin{aligned}
&(N_C,R_C)_I+(n,0)_I&,\quad&\text{for odd $n$,} \\
&(N_C,R_C)_I+(n,n/2)_I&,\quad&\text{for even $n$.}
\end{aligned}
\right.
\eea
For $n=2$, based on the equivalence relation \eqref{equivalence relation for interacting rotation} and the fact that the generator of the $\mathbb{Z}_4$ classification is $(1,0)_I$, the class of a system with $(N_C,R_C)_I$ is
\bea
[N_C-R_C \,\text{mod}\, 4]
\eea
For odd $n$, the $\mathbb{Z}_n\times\mathbb{Z}_n$ classification can be generated by $\{(1,0)_I,(0,1)_I\}$. Consequently, the equivalence relation \eqref{equivalence relation for interacting rotation} indicates that a system with $(N_C,R_C)_I$ belongs to the class
\bea
[(N_C\,\text{mod}\, n,R_C\,\text{mod}\, n)]
\eea
For even $n$ where $n\ge 4$, the equivalence relation \eqref{equivalence relation for interacting rotation} combined with the fact that the $\mathbb{Z}_{2n}\times\mathbb{Z}_{n/2}$ classification is generated by $\{(1,0)_I,(2,1)_I\}$ asserts that a system with $(N_C,R_C)_I$ falls into the class
\bea
\text{$[(A\,\text{mod}\, 2n,B\,\text{mod}\, n/2)]$ with $(N_C,R_C)_I=A(1,0)_I+B(2,1)_I$.}
\eea
Again, for uniformity, we assign the quantum number $(N_{CI}\in\mathbb{Z},N_{BIQH}\in\mathbb{Z})_I$ to describe the $\mathbb{Z}^2$ classification of $E^{1}_{2,-2}$, as this classification represents Chern insulators and Bosonic integer quantum Hall states in the 2-cell. With this expression, the quantum number $(N_{CI},N_{BIQH})_I$ corresponds to the class $[(N_{CI},N_{BIQH})]$. In summary, a system with the quantum numbers $(N_C,R_C)_I$ and $(N_{CI},N_{BIQH})_I$ falls into 
\bea
\begin{aligned}
&[N_C-R_C \,\text{mod}\, 4]\oplus[(N_{CI},N_{BIQH})]&,\quad&\text{for $n=2$,} \\
&[(N_C\,\text{mod}\, n,R_C\,\text{mod}\, n)]\oplus[(N_{CI},N_{BIQH})]&,\quad&\text{for odd $n$,} \\
&[(A\,\text{mod}\, 2n,B\,\text{mod}\, n/2)]\oplus[(N_{CI},N_{BIQH})]&,\quad&\text{for even $n$ and $n\ge 4$,}
\end{aligned}
\eea
where $A$ and $B$ are determined by $(N_C,R_C)_I=A(1,0)_I+B(2,1)_I$.

\subsection{Interaction effects on free-fermion crystalline SPT phases}
\begin{table}[htb!]
\centering
\renewcommand{\arraystretch}{2}
\begin{tabular}{c|c|c}
  Symmetry group  & Free-fermion SPT phases & Interacting SPT phases  \\ \hline 
$U(1)\times C_2$ & $[N_1]\oplus[N_{CI}]$ & $[-N_1\,\text{mod}\,4]\oplus[(N_{CI},0)]$ \\ \hline
\makecell{$U(1)\times C_n$ \\$n$ is odd} & $[(N_1,N_2,\ldots,N_{n-1})]\oplus[N_{CI}]$ & \makecell{$\left[(\sum_{\gamma=0}^{n-1}N_{\gamma}\,\,\text{mod}\,n,\sum_{\gamma=1}^{n-1}\gamma N_{\gamma}\,\,\text{mod}\,n)\right]$\\$\oplus[(N_{CI},0)]$} \\ \hline
\makecell{$U(1)\times C_n$ \\$n$ is even and $n\ge 4$} & $[(N_1,N_2,\ldots,N_{n-1})]\oplus[N_{CI}]$ & $\left[(A\,\,\text{mod}\,2n,B\,\,\text{mod}\,n/2)\right]\oplus[(N_{CI},0)]$ \\
\end{tabular}
\caption{The relation between free-fermion and interacting SPT phases for the $2d$ systems with $U(1)$ and $C_n$ rotation symmetry. Here, $N_{\gamma}=n_{\gamma}-n_0$, where $n_{\gamma}$ is defined in Sec.~\ref{SPT free roation}. The coefficients $A$ and $B$ are determined by $(\sum_{\gamma=0}^{n-1}N_{\gamma},\sum_{\gamma=1}^{n-1}\gamma N_{\gamma}\,\,\text{mod}\,n)_I =A(1,0)_I+B(2,1)_I$. For the symmetry group $U(1)\times C_2$, if we define $N_1=n_0-n_1$, then the reduction of $[N_1]$ will be $[N_1\,\text{mod}\,4]$, which parallels the reduction of $1d$ systems with $U(1)$ and reflection symmetry.}
\label{Interaction effects for 2d rotation}
\end{table}
The interaction effect on $2d$ free-fermion systems with $U(1)$ and $C_n$ rotation symmetry can be analyzed separately. Let's first focus on the free-fermion classification related to the SPT phases on the $2$-cell $\alpha$, $E^{1}_{2,-2}$. This classification refers to Chern insulators. Since interactions do not alter the classification of Chern insulators and there is no free-fermion description for Bosonic integer quantum Hall states, it is clear that interactions only turn the class $[N_{CI}]$ into $[(N_{CI},0)]$. In other words, we have
\bea\label{reduction of SPT on 2-cell for rotation}
[N_{CI}]\xrightarrow{\text{Interaction effect}}[(N_{CI},0)]
\eea

For understanding how interactions affect the classification pertainning to the SPT phases on the rotation center, we label the complex fermion with the rotation eigenvalue $e^{\gamma \frac{2\pi i}{n}}$ as $\phi^{\dag}_{\gamma}$. Since, as mentioned in Sec.~\ref{free to interaction, reflection}, a single-particle state can be regarded as a complex fermion, a free-fermion system falling into $[(N_1,N_2,\ldots,N_{n-1})]$ in the Fock space can be represented by the SPT state (in the 0-cell)
\bea
\ket{\psi}=\prod^{n-1}_{\gamma=0}\prod^{n_{\gamma}}_{j=1} \phi^{\dag}_{\gamma,j} \ket{0},
\eea
where $N_{\gamma}=n_{\gamma}-n_0$ as defined in Sec.~\ref{SPT free roation}, and one can consider $n_\gamma$ as the number of the complex fermion $\phi^{\dag}_{\gamma}$. Given that $C_{n}\phi^{\dag}_{\gamma,j}C_{n}^{-1}=e^{\gamma \frac{2\pi i}{n}}\phi^{\dag}_{\gamma,j}$ and using eq.~\eqref{NCRC rotation def}, the corresponding interacting quantum number is
\bea\label{interacting quantum number for free rotation}
(\sum_{\gamma=0}^{n-1}n_{\gamma},\sum_{\gamma=1}^{n-1}\gamma n_{\gamma}\,\,\text{mod}\,n)_I
\eea
For $n=2$, the relation~\eqref{equivalence relation for interacting rotation} simplifies the quantum number~\eqref{interacting quantum number for free rotation} to
\bea
\begin{aligned}
(n_{1}+n_{0},n_{1}\,\,\text{mod}\,2)_I&\sim(n_{1}+n_{0},n_{1}\,\,\text{mod}\,2)_I-n_{1}(2,1)_I=(n_{0}-n_{1},0)_I\\
&=(-N_1,0)_I\sim(-N_1\,\text{mod}\,4,0)_I.
\end{aligned}
\eea
Since the $\mathbb{Z}_4$ classification can be generated by $(1,0)_I$, with eq.~\eqref{reduction of SPT on 2-cell for rotation}, we have
\bea
[N_1]\oplus[N_{CI}]\xrightarrow{\text{Interaction effect}}[-N_1\,\text{mod}\,4]\oplus[(N_{CI},0)].
\eea
For odd $n$, applying the relation~\eqref{equivalence relation for interacting rotation} can transform the quantum number~\eqref{interacting quantum number for free rotation} into
\bea
\begin{aligned}
(\sum_{\gamma=0}^{n-1}n_{\gamma},\sum_{\gamma=1}^{n-1}\gamma n_{\gamma}\,\,\text{mod}\,n)_I&\sim(\sum_{\gamma=0}^{n-1}n_{\gamma},\sum_{\gamma=1}^{n-1}\gamma n_{\gamma}\,\,\text{mod}\,n)_I-n_0(n,0)_I\\
&=(\sum_{\gamma=0}^{n-1}(n_{\gamma}-n_0),\sum_{\gamma=1}^{n-1}\gamma n_{\gamma}\,\,\text{mod}\,n)_I\\
&=(\sum_{\gamma=0}^{n-1}(n_{\gamma}-n_0),\sum_{\gamma=1}^{n-1}\gamma n_{\gamma}\,\,\text{mod}\,n)_I-(0,\frac{n(n-1)}{2} n_0\,\,\text{mod}\,n)\\
&=(\sum_{\gamma=0}^{n-1}(n_{\gamma}-n_0),\sum_{\gamma=1}^{n-1}\gamma (n_{\gamma}-n_0)\,\,\text{mod}\,n)_I\\
&=(\sum_{\gamma=0}^{n-1}N_{\gamma},\sum_{\gamma=1}^{n-1}\gamma N_{\gamma}\,\,\text{mod}\,n)_I.
\end{aligned}
\eea
Here, $(0,0)_I=(0,\frac{n(n-1)}{2} n_0\,\,\text{mod}\,n)_I$ is given that $n$ is odd. With eq.~\eqref{reduction of SPT on 2-cell for rotation} and the fact that the $\mathbb{Z}_n\times\mathbb{Z}_n$ classification can be generated by $\{(1,0)_I,(0,1)_I\}$, the reduction of the class is 
\bea
[(N_1,N_2,\ldots,N_{n-1})]\oplus[N_{CI}]\xrightarrow{\text{Interaction effect}}\left[(\sum_{\gamma=0}^{n-1}N_{\gamma}\,\,\text{mod}\,n,\sum_{\gamma=1}^{n-1}\gamma N_{\gamma}\,\,\text{mod}\,n)\right]\oplus[(N_{CI},0)].
\eea
For even $n$ and $n\ge 4$, the relation~\eqref{equivalence relation for interacting rotation} allows us to rewrite the quantum number~\eqref{interacting quantum number for free rotation} as
\bea
\begin{aligned}
(\sum_{\gamma=0}^{n-1}n_{\gamma},\sum_{\gamma=1}^{n-1}\gamma n_{\gamma}\,\,\text{mod}\,n)_I&\sim(\sum_{\gamma=0}^{n-1}n_{\gamma},\sum_{\gamma=1}^{n-1}\gamma n_{\gamma}\,\,\text{mod}\,n)_I-n_0(n,n/2)_I\\
&=(\sum_{\gamma=0}^{n-1}(n_{\gamma}-n_0),\sum_{\gamma=1}^{n-1}\gamma n_{\gamma}\,\,\text{mod}\,n)_I-(0,n_0\frac{n}{2}\,\,\text{mod}\,n)_I\\
\end{aligned}
\eea
Because $\frac{n}{2}\,\text{mod}\,n=-\frac{n}{2}\,\text{mod}\,n$, we have $(0,n_0\frac{n}{2}\,\,\text{mod}\,n)_I=-(0,n_0\frac{n}{2}\,\,\text{mod}\,n)_I$ and
\bea
\begin{aligned}
(\sum_{\gamma=0}^{n-1}n_{\gamma},\sum_{\gamma=1}^{n-1}\gamma n_{\gamma}\,\,\text{mod}\,n)_I&\sim(\sum_{\gamma=0}^{n-1}(n_{\gamma}-n_0),\sum_{\gamma=1}^{n-1}\gamma n_{\gamma}\,\,\text{mod}\,n)_I+(0,n_0\frac{n}{2}\,\,\text{mod}\,n)_I\\
&\begin{aligned}
=&(\sum_{\gamma=0}^{n-1}(n_{\gamma}-n_0),\sum_{\gamma=1}^{n-1}\gamma n_{\gamma}\,\,\text{mod}\,n)_I+(0,n_0\frac{n}{2}\,\,\text{mod}\,n)_I\\
&-(0,n_0\frac{n^2}{2}\,\,\text{mod}\,n)_I
\end{aligned}
\\
&=(\sum_{\gamma=0}^{n-1}(n_{\gamma}-n_0),\sum_{\gamma=1}^{n-1}\gamma n_{\gamma}\,\,\text{mod}\,n)_I-(0,\frac{n(n-1)}{2} n_0\,\,\text{mod}\,n)\\
&=(\sum_{\gamma=0}^{n-1}(n_{\gamma}-n_0),\sum_{\gamma=1}^{n-1}\gamma (n_{\gamma}-n_0)\,\,\text{mod}\,n)_I\\
&=(\sum_{\gamma=0}^{n-1}N_{\gamma},\sum_{\gamma=1}^{n-1}\gamma N_{\gamma}\,\,\text{mod}\,n)_I.
\end{aligned}
\eea
Here, $(0,0)_I=(0,n_0\frac{n^2}{2}\,\,\text{mod}\,n)_I$ since $n$ is even. By combining eq.~\eqref{reduction of SPT on 2-cell for rotation} and the fact that the generators of the $\mathbb{Z}_{2n}\times\mathbb{Z}_{n/2}$ classification are $(1,0)_I$ and $(2,1)_I$, we obtain
\bea
[(N_1,N_2,\ldots,N_{n-1})]\oplus[N_{CI}]\xrightarrow{\text{Interaction effect}}[(A\,\text{mod}\, 2n,B\,\text{mod}\, n/2)]\oplus[(N_{CI},0)],
\eea
where $(\sum_{\gamma=0}^{n-1}N_{\gamma},\sum_{\gamma=1}^{n-1}\gamma N_{\gamma}\,\,\text{mod}\,n)_I =A(1,0)_I+B(2,1)_I$. We collect the results here in Table~\ref{Interaction effects for 2d rotation}.
\section{\texorpdfstring{3$d$}{\textmu} systems with \texorpdfstring{$U(1)$}{\textmu} and inversion symmetry}\label{Main inversion Sec}
\begin{figure}[htb!]
\centering
\includegraphics[width=0.4\textwidth]{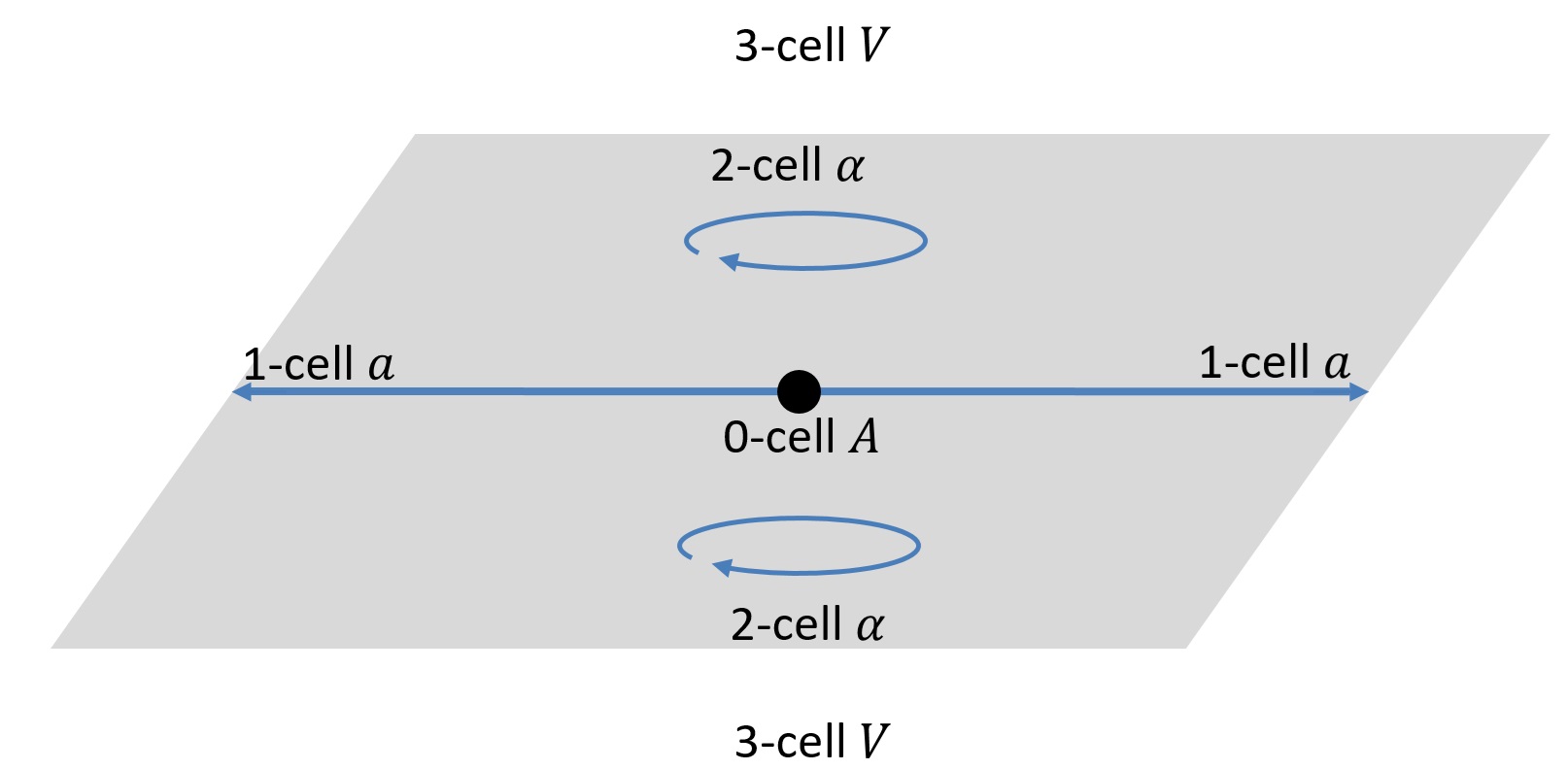}
\caption{The $\mathbb{Z}_2$-symmetric cell decomposition of $3d$ space. The 0-cell $A$ has $U(1)$ and inversion symmetry, while the 1-cell $a$, the 2-cell $\alpha$, and the 3-cell $V$ have $U(1)$ symmetry only.}
\label{cell_decomposition_inversion}
\end{figure}
In this section, we demonstrate how our approach can be applied to investigate the interaction effects on crystalline SPT phases in cases where the classification fits into non-split short exact sequences. As an example, we consider 3$d$ systems with $U(1)$ and inversion symmetry. Their free-fermion and interacting classification are given by the $K$-homology $K_{0}^{\mathbb{Z}_2}(\mathbb{R}^3,\partial \mathbb{R}^3)$ and the generalized homology $h_{0}^{\mathbb{Z}_2}(\mathbb{R}^3,\partial \mathbb{R}^3)$, respectively. With eq.~\eqref{short exact sequences}, we have
\bea\label{short exact sequence for 3d inversion}
\begin{alignedat}{3}
0\rightarrow &  E^{4}_{0,0}&&\rightarrow \quad\quad\quad\quad\,F_{1}h_{0}&&\rightarrow E^{4}_{1,-1}\rightarrow 0,\\
0\rightarrow & F_{1}h_{0}&&\rightarrow \quad\quad\quad\quad\,F_{2}h_{0}&&\rightarrow E^{4}_{2,-2}\rightarrow 0,\\
0\rightarrow &  F_{2}h_{0}&&\rightarrow \text{$K_{0}^{\mathbb{Z}_2}(\mathbb{R}^3,\partial \mathbb{R}^3)$ or $ h_{0}^{\mathbb{Z}_2}(\mathbb{R}^3,\partial \mathbb{R}^3)$}&&\rightarrow E^{4}_{3,-3}\rightarrow 0.\\
\end{alignedat}
\eea
We will show that this set of short exact sequences can be written as
\bea
0\rightarrow E^{2}_{0,0}\rightarrow \text{$K_{0}^{\mathbb{Z}_2}(\mathbb{R}^3,\partial \mathbb{R}^3)$ or $ h_{0}^{\mathbb{Z}_2}(\mathbb{R}^3,\partial \mathbb{R}^3)$}\rightarrow E^{2}_{2,-2}\rightarrow 0.
\eea
The trivial solution to this sequence is $ \text{$K_{0}^{\mathbb{Z}_2}(\mathbb{R}^3,\partial \mathbb{R}^3)$ or $ h_{0}^{\mathbb{Z}_2}(\mathbb{R}^3,\partial \mathbb{R}^3)$}\cong E^{2}_{0,0}\oplus E^{2}_{2,-2}$. However, in this case,
because the trivial phase of $E^{2}_{2,-2}$ can generate $E^{2}_{0,0}$, the sequence should be treated as a non-split short exact sequence, leading to a non-trivial solution. More details will be included in the following discussion.
\subsection{Free-fermion crystalline SPT phases}\label{SPT free inversion}
\begin{figure}[htp!]
\centering
\subfloat[]{\includegraphics[width=0.35\textwidth]{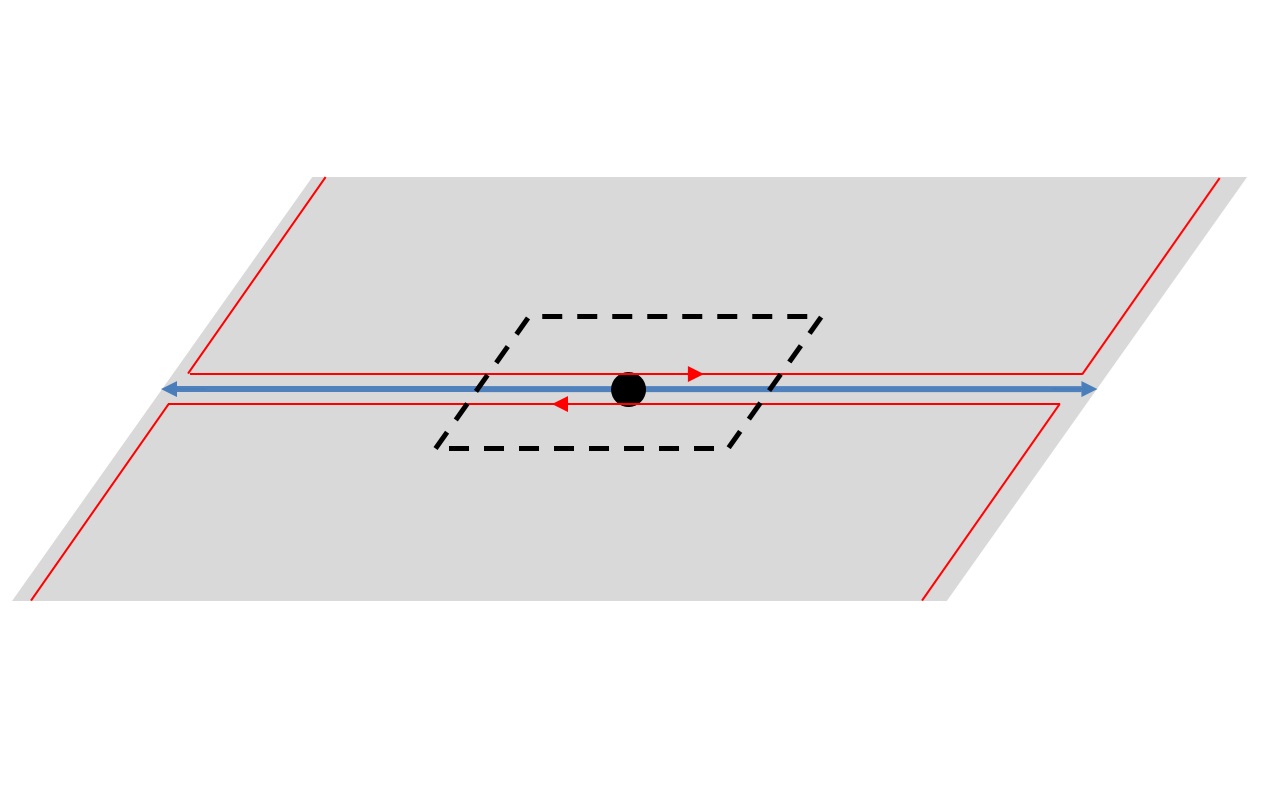}\label{boundary_cancel}}\hskip 0.5cm
\subfloat[]{\includegraphics[width=0.35\textwidth]{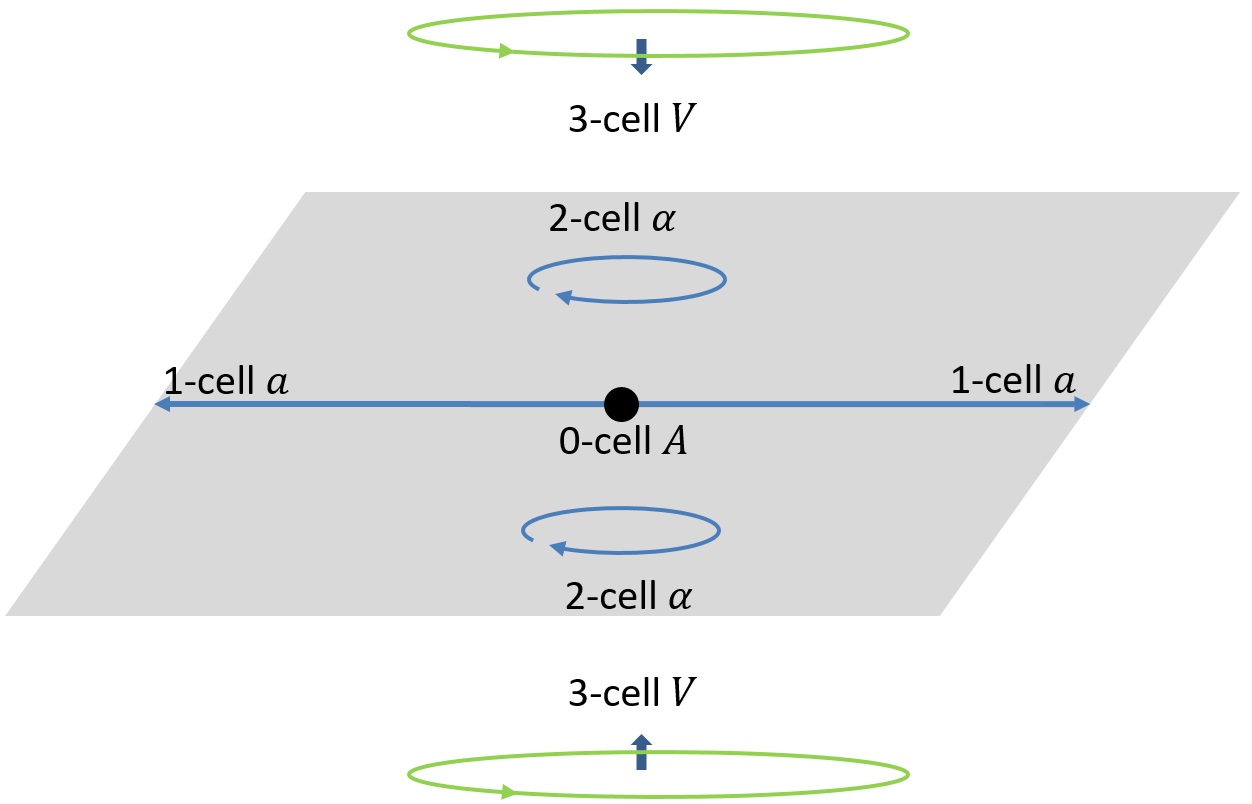}\label{First_differential_inversion}}\\
\caption{(a) An inversion-symmetric pair of chiral edge states cancels each other out. (b) The first differential $d^{1}_{3,2}$ (which is equivalent to $d^{1}_{3,0}$ in the free-fermion case) under the inversion symmetry.}
\end{figure}
Following the $\mathbb{Z}_2$-symmetric cell decomposition shown in Fig.~\ref{cell_decomposition_inversion}, the $E^1$-page for 3$d$ free-fermion systems with $U(1)$ and inversion symmetry is given by~\cite{freeAHSSinrealspace}
\bea\label{E-1 page, 3d free inversion}
\centering
\renewcommand{\arraystretch}{1.5}
\begin{tabular}{c|cccc}
$q=0$        &  $\mathbb{Z}^2$ & $\mathbb{Z}$ & $\mathbb{Z}$ & $\mathbb{Z}$\\
$q=1$        & $0$           & $0$ & $0$ & $0$\\ \hline
$E^{1}_{p,-q}$ & $p=0$         & $p=1$ & $p=2$ & $p=3$   \\
\end{tabular}.
\eea
Because $E^{1}_{0,0}$ counts the number of SPT states in the irreps with inversion eigenvalue $+1$ and $-1$, we can use the quantum number $(n_+,n_-)_f$ to describe this $\mathbb{Z}^2$ classification. Here, $n_+$ and $n_-$ correspond to the number of SPT states with inversion eigenvalues $+1$ and $-1$. The terms $E^{1}_{p\in\{1,3\},0}$ are related to SPT states within Class AIII in the $p$-cell, which can be characterized by the quantum number $(W_p)_f$, where $W_p$ is the $p$-dimensional winding number. The element $E^{1}_{2,0}$ represents the Chern number $N_{CI}$ of Chern insulators in the 2-cell, so we assign the quantum number $(N_{CI})_f$ to it. With these notations, the first differentials can be expressed as
\bea
\begin{aligned}
&d^{1}_{1,0}:1\rightarrow(n_+,n_-)_f=(1,1)_f,\\
&d^{1}_{2,0}:1\rightarrow(W_p)_f=(0)_f,\\
&d^{1}_{3,0}:1\rightarrow(N_{CI})_f=(2)_f.\\
\end{aligned}
\eea
$d^{1}_{1,0}$ describes the pumping process illustrated in Fig.~\ref{First_differential_reflection}. The differential $d^{1}_{2,0}$ is trivial because inversion does not change the Hall conductivity. This trivial differential can be understood as the cancellation of boundary anomalies of SPT phases in the 2-cell, as shown in Fig.~\ref{boundary_cancel}. The differential $d^{1}_{3,0}$ represents the process of pumping two SPT states from the north and south 3-cells onto the 2-cell, as depicted in Fig.~\ref{First_differential_inversion}. Taking the homology of the first differentials, we get the $E^2$-page
\bea\label{E-2 page, 3d free inversion}
\centering
\renewcommand{\arraystretch}{1.5}
\begin{tabular}{c|cccc}
$q=0$        &  $\mathbb{Z}$ & $0$ & $\mathbb{Z}_2$ & $0$\\
$q=1$        & $0$           & $0$ & $0$ & $0$\\ \hline
$E^{1}_{p,-q}$ & $p=0$         & $p=1$ & $p=2$ & $p=3$   \\
\end{tabular}.
\eea
The $\mathbb{Z}$ classification of $E^{2}_{0,0}$ can be expressed in terms of $(n_+,n_-)_f$ with the equivalence relation
\bea
(n_+,n_-)_f+(1,+1)_f\sim(n_+,n_-)_f.
\eea
Therefore, this $\mathbb{Z}$ classification can be generated by $(1,0)_f$ or $(0,1)_f$. For $E^{2}_{2,0}$, the $\mathbb{Z}_2$ classification can be represented by $(N_{CI})_f$ with the equivalence relation
\bea
(N_{CI})_f+(2)_f\sim(N_{CI})_f.
\eea
Hence, it is generated by $(1)_f$.
\begin{figure}[htb!]
\centering
\includegraphics[width=0.4\textwidth]{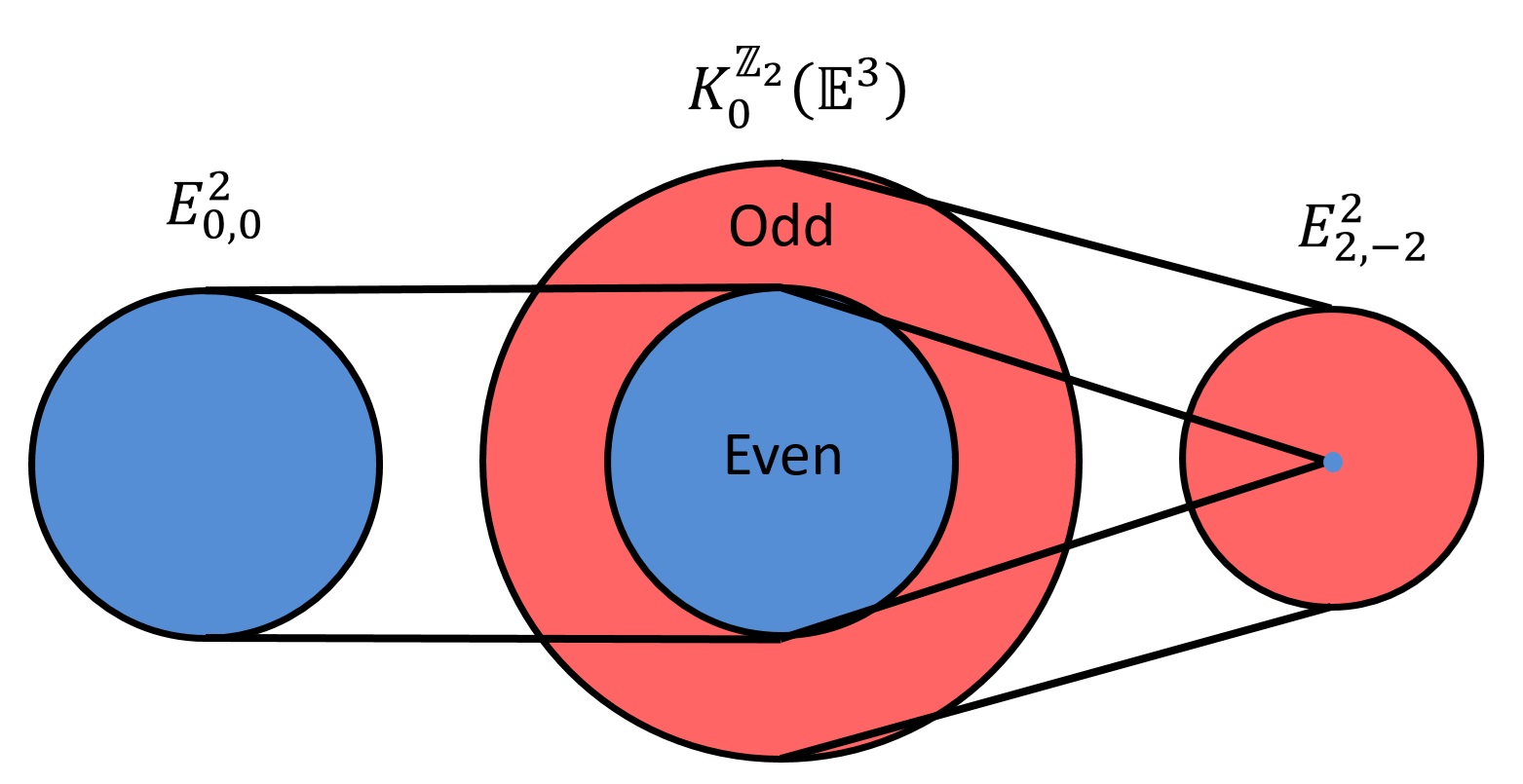}
\caption{Schematic depiction of the relation among $E^{2}_{0,0}$, $E^{2}_{2,0}$, and $K_{0}^{\mathbb{Z}_2}(\mathbb{R}^3,\partial \mathbb{R}^3)$. The labels "Even" and "Odd" represent even and odd values of the quantum number $N_{2-cell}$, respectively.}
\label{nontrivial_group_extension}
\end{figure}

\begin{figure}[htb!]
\centering
\includegraphics[width=0.6\textwidth]{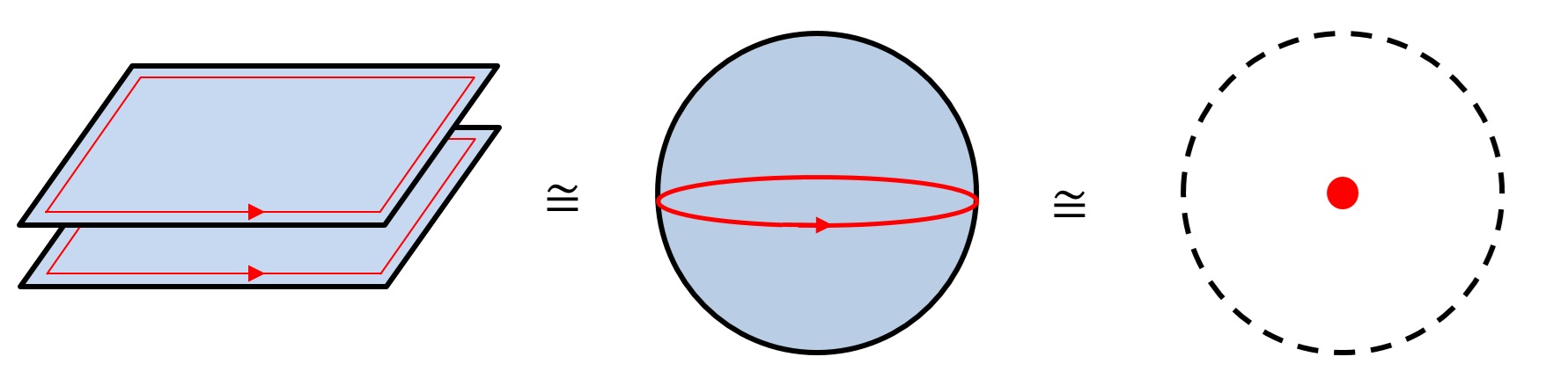}
\caption{Two layers of Chern insulators can be deformed into a monopole charge with even parity at the inversion center.}
\label{Chern_to_charge}
\end{figure}

Since there are no higher differentials, the $E^2$-page is the same as the $E^3$- and $E^4$-page. Plugging these results into the short exact sequences~\eqref{short exact sequence for 3d inversion}, we obtain
\bea
0\rightarrow E^{2}_{0,0}=\mathbb{Z}\rightarrow K_{0}^{\mathbb{Z}_2}(\mathbb{R}^3,\partial \mathbb{R}^3)\rightarrow E^{2}_{2,-2}=E^{2}_{2,0}=\mathbb{Z}_2\rightarrow 0.
\eea
Here, we should treat the sequence as a non-split short exact sequence, as $E^{2}_{0,0}$ can be generated by the trivial phase of $E^{2}_{2,0}$. To illustrate the non-split short exact sequence more clearly, we present the relation among $E^{2}_{0,0}$, $E^{2}_{2,0}$, and $K_{0}^{\mathbb{Z}_2}(\mathbb{R}^3,\partial \mathbb{R}^3)$ in Fig.~\ref{nontrivial_group_extension}. This generation can be explained as follows: as shown in Fig.~\ref{Chern_to_charge}, two layers of Chern insulators can be smoothly deformed into a Chern insulator defined on $S^2$ with the periodic boundary condition along $S^1$. Such a system is topologically equivalent to a monopole charge with even parity at the inversion center. This equivalence can be verified by analyzing the corresponding continuous Dirac Hamiltonian with a texture of a mass term~\cite{interactingAHSS,freeAHSSinrealspace}. Consequently, two layers of Chern insulators in the 2-cell, characterized by the quantum number $(N_{CI}=2)_f$, can be deformed into an SPT state in the 0-cell with $(n_+=1,n_-=0)_f$ through the differentials $d^{1}_{2,-2}$ and $d^{2}_{2,-2}$, where $(n_+=1,n_-=0)_f$ is the generator of $E^{2}_{0,0}$. In other words, we have $(N_{CI}=2)_f\cong(n_+=1,n_-=0)_f$, indicating that $(N_{CI}=2)_f$ can serve as the generator of $E^{2}_{0,0}$. Thus, the non-trivial solution to this short exact sequence is
\bea
 K_{0}^{\mathbb{Z}_2}(\mathbb{R}^3,\partial \mathbb{R}^3)=\mathbb{Z},
\eea
and this $\mathbb{Z}$ classification can be characterized by the quantum number $(N_{2-cell})_f=2(n_{+}-n_{-})_{f}+(N_{CI})_f$, where $2(n_{+}-n_{-})_f$ results from deforming the SPT state with $(n_+,n_-)_f$ at the inversion center to the SPT state with $(N_{CI}=2n_{+}-2n_{-})_f$ in the 2-cell. Thus, a free-fermion system with $(N_{2-cell})_f$ falls into the class
\bea
[N_{2-cell}].
\eea
More precisely, a system with the quantum number $(N_{2-cell})_f$ can be viewed as having a trivial phase on the 0-cell and a topological phase on the 2-cell, characterized by the Chern number $N_{2-cell}$—that is, as a system with $(n_+=0,n_-=0)_f$ and $(N_{CI}=N_{2-cell})_f$. Such a system can be deformed into one with quantum numbers $(n_+,n_-)_f$ and $(N_{CI})_f$, implying that the Chern number of Chern insulators in 2-cell (together with a trivial state in the 0-cell) is sufficient to describe the classification $K_{0}^{\mathbb{Z}_2}(\mathbb{R}^3,\partial \mathbb{R}^3)$.
\subsection{Interacting crystalline SPT phases}
The $E^1$-page of $3d$ interacting systems with $U(1)$ and inversion symmetry is given by~\cite{interactingAHSS}
\bea\label{E-1 page, 3d interacting inversion}
\centering
\renewcommand{\arraystretch}{1.5}
\begin{tabular}{c|cccc}
$q=0$        &  $\mathbb{Z}\times \mathbb{Z}_2$ & $\mathbb{Z}$ & $\mathbb{Z}$ & $\mathbb{Z}$\\
$q=1$        & $0$           & $0$ & $0$ & $0$\\ 
$q=2$        & $0$           & $\mathbb{Z}^2$ & $\mathbb{Z}^2$ & $\mathbb{Z}^2$\\ 
$q=3$        &            & $0$ & $0$ & $0$\\ \hline
$E^{1}_{p,-q}$ & $p=0$         & $p=1$ & $p=2$ & $p=3$  \\
\end{tabular}.
\eea
$E^{1}_{0,0}$ represents the integer-valued $U(1)$ charge and the inversion eigenvalue of a SPT state $\ket{\psi} = \psi^{\dag}\ket{0}$, which can be described by the quantum number $(N_C,I_C)_I$, where
\bea\label{NCRC inversion def}
N_C=\bra{\psi}\hat{N}\ket{\psi}\in \mathbb{Z}, \quad I_C=\left\{ 
\begin{aligned}
0,\quad&\text{if} \ I\psi^{\dag} I^{-1}=\psi,&\\
1,\quad&\text{if} \ I\psi^{\dag} I^{-1}=-\psi,&
\end{aligned}
\in \mathbb{Z}_2.
\right.
\eea
$I$ denotes the inversion operator. The $\mathbb{Z}^2$ classification of $E^{1}_{p\in\{1,2,3\},-2}$ corresponds to Chern insulators and Bosonic integer quantum Hall states in the 2-cell, so we assign the quantum number $(N_{CI},N_{BIQH})_I$ to this classification. In terms of these quantum numbers, the first differentials can be written as
\bea\label{first differential for 3d interacting inversion}
\begin{aligned}
&d^{1}_{1,0}:1\rightarrow(N_C,I_C)_I=(2,1)_I,\\
&d^{1}_{2,-2}:(n,m)\rightarrow(N_{CI},N_{BIQH})_I=(0,0)_I,\\
&d^{1}_{3,-2}:(n,m)\rightarrow(N_{CI},N_{BIQH})_I=(2n,2m)_I.\\
\end{aligned}
\eea
$d^{1}_{1,0}$ is about a pumping process similar to the one when discussing $d^{1}_{1,0}$ for 1$d$ interacting system with $U(1)$ and reflection symmetry. The differential $d^{1}_{2,2}$ is trivial because inversion does not affect the Hall conductivity and the thermal Hall conductivity. $d^{1}_{3,2}$ illustrates the process of pumping SPT states from the 3-cell onto the 2-cell while preserving inversion symmetry, as shown in Fig.~\ref{First_differential_inversion}. The homology of these first differentials leads to the $E^2$-page
\bea\label{$E-2$ page, 3d interacting inversion}
\centering
\renewcommand{\arraystretch}{1.5}
\begin{tabular}{c|cccc}
$q=0$        &  $\mathbb{Z}_4$ & $0$ &  & \\
$q=1$        & $0$           & $0$ & $0$ & $0$\\ 
$q=2$        &            &  & $\mathbb{Z}_2\times\mathbb{Z}_2$ & $0$\\ 
$q=3$        &            & $0$ & $0$ & $0$\\ \hline
$E^{1}_{p,-q}$ & $p=0$         & $p=1$ & $p=2$ & $p=3$  \\
\end{tabular}.
\eea
The absence of higher differentials ensures that the $E^2$-page is the same as the $E^4$-page. Consequently, the short exact sequences~\eqref{short exact sequence for 3d inversion} can be written as
\bea\label{short exact sequence for 3d inversion interacting}
0\rightarrow E^{2}_{0,0}=\mathbb{Z}_4\rightarrow  h_{0}^{\mathbb{Z}_2}(\mathbb{R}^3,\partial \mathbb{R}^3)\rightarrow E^{2}_{2,-2}=\mathbb{Z}_2\times\mathbb{Z}_2\rightarrow 0.
\eea
We need to account for the non-trivial solution to the short exact sequence, as the trivial phase of $E^{2}_{2,-2}$ generates $E^{2}_{0,0}$. As discussed in Sec.~\ref{SPT free inversion}, two layers of Chern insulators in the 2-cell are topologically equivalent to a monopole charge with even parity at the inversion center, which results in
\bea\label{chern to charge quantum number}
(N_{CI}=2,N_{BIQH})_I\cong(N_C=1,I_C=0)_I\oplus(N_{BIQH})_I.
\eea
Note that $E^{2}_{0,0}$ can be generated by $(N_C=1,I_C=0)_I$, so $(N_{CI}=2)_I$, which is the trivial phase of $E^{2}_{2,-2}$, can serve as the generator of $E^{2}_{0,0}$. The group extension associated with bosonic integer quantum Hall states is trivial because double-layer bosonic integer quantum Hall states cannot be represented by a monopole charge in the same way as double-layer Chern insulators. Thus, we obtain
\bea
h_{0}^{\mathbb{Z}_2}(\mathbb{R}^3,\partial \mathbb{R}^3)=\mathbb{Z}_8\times\mathbb{Z}_2.
\eea
This classification can be described by the quantum number $(N_{2-cell},N_{BIQH})_I$, with the equivalence relations from the first differentials~\eqref{first differential for 3d interacting inversion}
\bea
\begin{aligned}
&(N_{2-cell},N_{BIQH})_I+(8,0)\sim(N_{2-cell},N_{BIQH})_I,\\
&(N_{2-cell},N_{BIQH})_I+(0,2)\sim(N_{2-cell},N_{BIQH})_I,
\end{aligned}
\eea
where $N_{2-cell}=N_{CI}+2(N_C-2I_C)$. The term $2(N_C-2I_C)$ in $N_{2-cell}$ is the Chern number of the $2d$ SPT state deformed from a $0d$ SPT state with $(N_C,I_C)_I$, which means that a system with $(N_{2-cell},N_{BIQH})_I$ can be viewed as one with a trivial state in the 0-cell and a $2d$ SPT state in the 2-cell, characterized by the Chern number $N_{2-cell}$ and the quantum number $N_{BIQH}$. The first equivalence relation arises from the fact that $E^{2}_{0,0}\cong E^{1}_{0,0}/\text{Im}(d^{1}_{1,0})$ gives the equivalence relation $(N_C,I_C)_I+(2,1)_I\sim (N_C,I_C)_I$, which leads to 
\bea\label{trivialization of chern to charge quantum number}
\begin{aligned}
(N_{2-cell}=8n)_I&\sim(N_C=4n+2I_C,I_C)_I\sim(4n+2I_C,I_C)_I-I_C(2,1)_I=(4n,0)_I\\
&\sim(4n,0)_I-2n(2,1)_I=(0,0)_I.
\end{aligned}
\eea
The relation $(N_{2-cell}=8n)_I\sim(4n+2I_C,I_C)_I$ is derived from deforming a system with $(N_{2-cell}=8n)_I$ into a system with $N_{CI}=0$. For the short exact sequence~\eqref{short exact sequence for 3d inversion interacting}, the map from $E^{2}_{0,0}$ to $h_{0}^{\mathbb{Z}_2}(\mathbb{R}^3,\partial \mathbb{R}^3)$ is a monomorphism. Therefore, combined with eq.~\eqref{trivialization of chern to charge quantum number}, we conclude that $(N_{2-cell}=8n)_I$ is in the trivial class. As a result, an interacting system with the quantum number $(N_{2-cell},N_{BIQH})_I$ falls into the class
\bea
[(N_{2-cell}\,\,\text{mod}\,8,N_{BIQH}\,\,\text{mod}\,2)].
\eea
Notably, the fact that the equivalence classes can be characterized by $(N_{2-cell},N_{BIQH})_I$ means that the classification here can be sufficiently described by the $2d$ SPT states in the 2-cell (together with a trivial state in the 0-cell), similar to its free-fermion counterpart.
\subsection{Interaction effects on free-fermion crystalline SPT phases}
\begin{table}[htb!]
\centering
\renewcommand{\arraystretch}{1.5}
\begin{tabular}{c|c|c}
  Symmetry group  & Free-fermion SPT phases & Interacting SPT phases  \\ \hline 
$U(1)\times \mathbb{Z}_2^{I}$ & $[N_{2-cell}]$ & $[(N_{2-cell}\,\text{mod}\,8,0)]$ \\ 
\end{tabular}
\caption{The relation between free-fermion and interacting SPT phases for $3d$ systems with $U(1)$ and inversion symmetry.}
\end{table}
The interaction effect on $3d$ free-fermion systems with $U(1)$ and inversion symmetry can be understood using the quantum number previously proposed. First, there is no free-fermion description for Bosonic integer quantum Hall states. Second, both the free-fermion classification and the $\mathbb{Z}_8$ classification of interacting systems can be described by the Chern number of Chern insulators in the 2-cell (together with a trivial state in the 0-cell). Considering these points and the fact that the Chern number remains invariant in the presence of interactions, a free-fermion system in the class $[N_{2-cell}]$ corresponds to an interacting system with the quantum number $(N_{2-cell},N_{BIQH}=0)_I$, namely that
\bea
[N_{2-cell}]\xrightarrow{\text{Interaction effect}}[(N_{2-cell}\,\text{mod}\,8,0)].
\eea

\section{Conclusion}
In this work, we have demonstrated that the AHSS provides a robust framework for analyzing the effects of interactions on crystalline SPT phases. Using the AHSS, the classification of crystalline SPT phases can be determined by solving a series of short exact sequences involving SPT phases protected by internal symmetries on subspaces. In other words, crystalline SPT phases can be effectively represented as SPT phases defined on these subspaces. This representation enables the assignment of specific symmetry quantum numbers to characterize crystalline SPT phases in both the free-fermion and interacting regimes. In particular, the reduction of free-fermion equivalence classes due to interactions is clearly elucidated through the connection between the symmetry quantum numbers in these two regimes. We apply this approach to fermionic systems with various symmetries in 1$d$, 2$d$, and 3$d$, including cases involving non-split short exact sequences.

%In this work, we have demonstrated that the AHSS provides a robust framework for analyzing interaction effects on crystalline SPT phases. By using the AHSS, the classification of crystalline SPT phases can be determined by solving a series of short exact sequences involving SPT phases protected by internal symmetries on subspaces. In other words, crystalline SPT phases can be effectively represented by SPT phases defined on these subspaces. This representation allows us to assign certain quantum numbers to describe crystalline SPT phases, whether in the free-fermion or interacting regime. By establishing a clear connection between free-fermion and interacting quantum numbers, we can discern how interactions modify the free-fermion equivalence classes. The examples discussed in this paper show that this approach is versatile and applicable across different dimensions, including cases with non-split short exact sequences.

\medskip
Acknowledgments---
C.-S. L. is supported by National Science and Technology Council Graduate Research Fellowship Pilot Program under Grant No. 113-2926-I-002-002-MY3.
K.S. is supported by JST CREST Grant No.~JPMJCR19T2 and JSPS KAKENHI Grant Nos.~22H05118 and 23H01097. 
C.-T. H. is supported by the Yushan (Young) Scholar Program under Grant No. NTU-111VV016 and by the National Science and Technology Council (NSTC) of Taiwan under Grant No. 112-2112-M-002-048-MY3.

\begin{appendices}
\section{\texorpdfstring{Decomposable systems and $1d$ SPT phases protected by $U(1)$ and reflection symmetry}{\textmu}}\label{Hpn and SPT phases}
In this section,  we first present how to assign the quantum numbers introduced in Sec.~\ref{1d R free SPT sec} and Sec.~\ref{1d R interacting SPT sec} to a specific type of system, which we refer to as decomposable systems. Next, we introduce a series of lattice models to illustrate the relation between free-fermion and interacting classes established in Sec.~\ref{free to interaction, reflection}.
\subsection{Decomposable systems and quantum numbers}\label{Decomposable systems and quantum numbers}
\begin{figure}[htp!]
\centering
\subfloat[]{\includegraphics[width=0.45\textwidth]{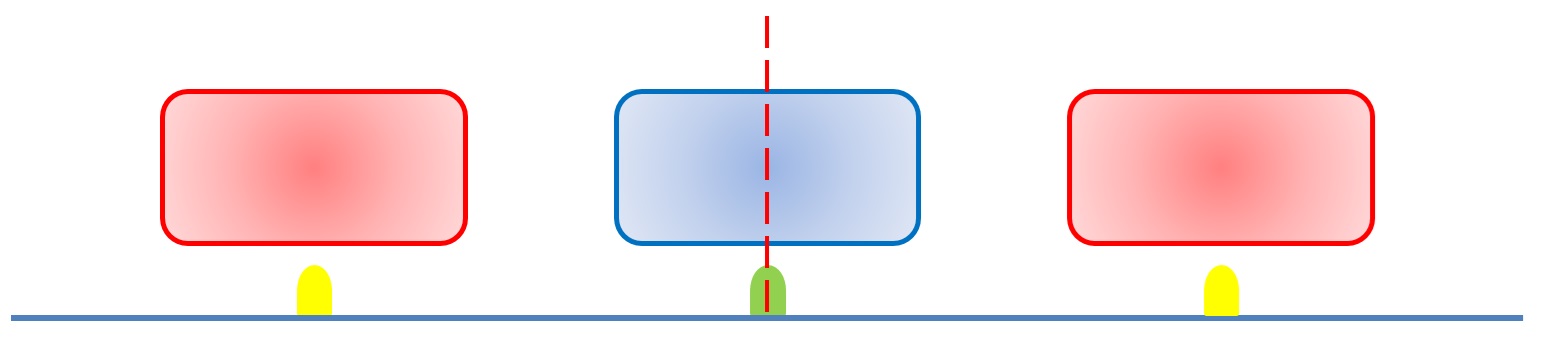}}\hskip 0.5cm
\subfloat[]{\includegraphics[width=0.45\textwidth]{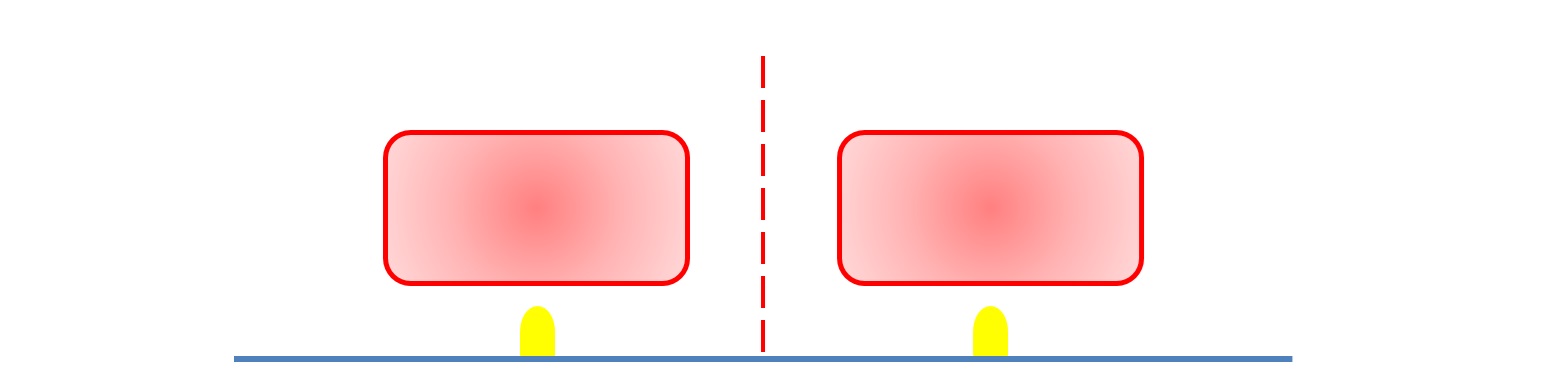}}\\
\caption{Schematic illustration of decomposable systems.The shadowed and rounded rectangles signify the subsystems of a decomposable system. The red dashed line marks the reflection center. The yellow and green wave packets represent charges defined for distinct subsystems. For decomposable systems with reflection symmetry, an appropriate choice of basis allows the reflection center to be positioned either at the midpoint of a single subsystem, as illustrated in (a), or between two subsystems, as shown in (b).}
\label{Decomposable systems}
\end{figure}
\begin{figure}[htp!]
\centering
\subfloat[]{\includegraphics[width=0.45\textwidth]{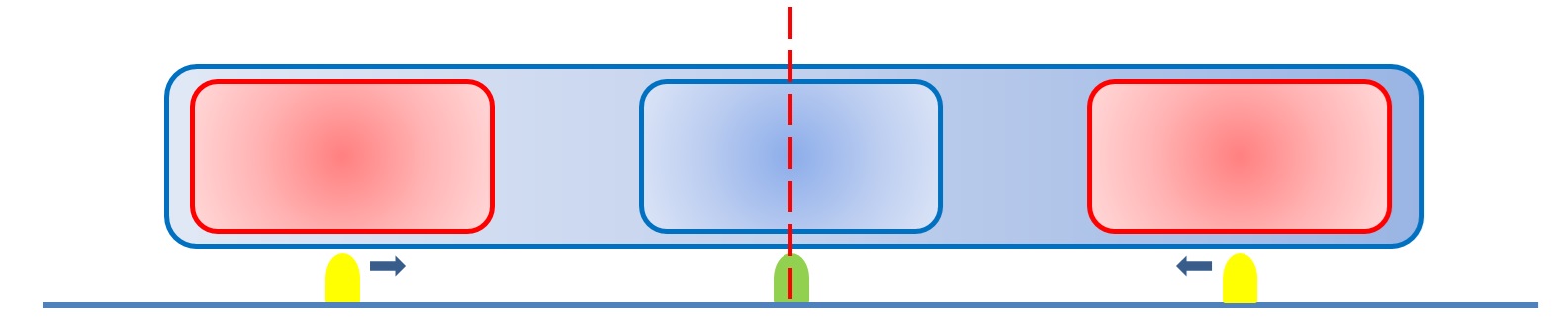}}\hskip 0.5cm
\subfloat[]{\includegraphics[width=0.45\textwidth]{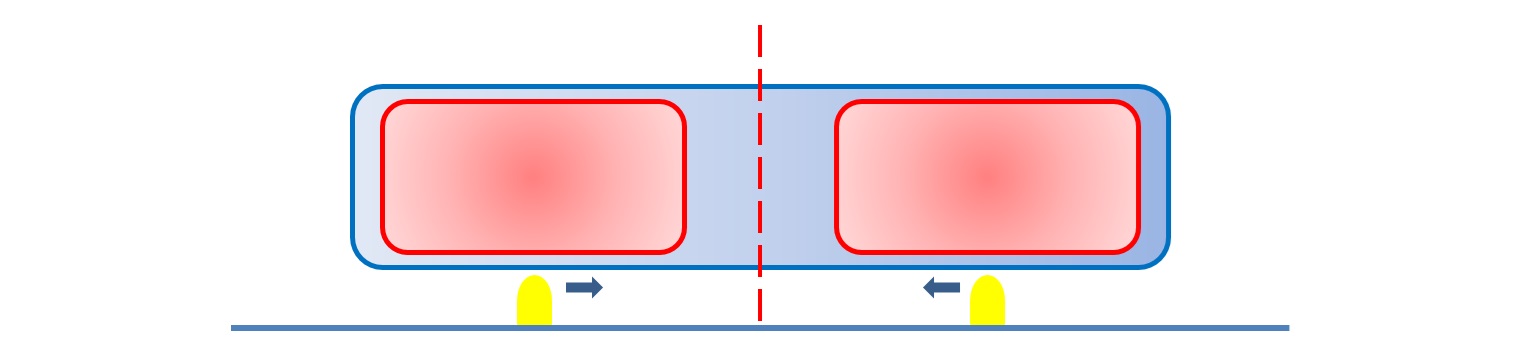}}\\
\caption{The first differential $d^{1}_{1,0}$, which represents a process of pumping SPT states in 1-cell onto 0-cell, can be realized by expanding the subsystem to include neighboring subsystems while maintaining reflection symmetry.}
\label{First differential and decomposable systems}
\end{figure}

In the work~\cite{Ourcrystallineequivalence}, the authors showed that the quantum number $(N_C,R_C)_I$ can be assigned to decomposable systems—systems that can be further divided into subsystems, where no hopping or interaction occurs between different subsystems, as illustrated in Fig.~\ref{Decomposable systems}. With this definition, the many-body ground state of a decomposable system can be expressed as
\bea
\ket{\psi} = \ket{\psi_1}\ket{\psi_2}\ket{\psi_3}\ldots=\prod_j \psi_{j}\ket{0},
\eea
where $j$ denotes the index of the subsystem. The many-body ground state of each subsystem, $\ket{\psi_n}$, can be regarded as a charge, with its location defined at the midpoint of the subsystem. Under this framework, if the location of a charge $\ket{\psi_C}$ coincides with the reflection center, $\ket{\psi_C}$ can be considered as the SPT state on the 0-cell. Consequently, the quantum number $(N_C,R_C)_I$ of the decomposable system can be determined by $\ket{\psi_C}$. Conversely, if there is no $\ket{\psi_j}$ located at the reflection center, the decomposable system is topologically trivial. Two important points should be noted. First, since the decomposable systems we discuss here respect reflection symmetry, an appropriate choice of basis allows the reflection center to either coincide with the midpoint of a subsystem or be located between two subsystems, as shown in Fig.~\ref{Decomposable systems}. Second, subsystems can be enlarged by incorporating neighboring subsystems while preserving reflection symmetry. This process physically realizes the first differential $d^{1}_{1,0}$ introduced earlier, as depicted in Fig.~\ref{First differential and decomposable systems}.

Since, in the free-fermion setting, the $0d$ SPT state $\ket{\psi_C}$ can be interpreted as multiple single-particle fermions (i.e., single-particle states), we can assign the quantum number $(n_{+},n_{-})_f$ for free-fermion decomposable systems in a similar way. Specifically, for a decomposable free-fermion system with translation symmetry, the unit cell can serve as the subsystem. If the reflection center is located between two unit cells, the decomposable system is trivial. However, when the reflection center is positioned at the midpoint of the unit cell after choosing an appropriate basis, we can assign the quantum number $(n_{+},n_{-})_f$ to the decomposable system, where $n_{\pm}$ is determined by the number of occupied eigenstates of the Bloch Hamiltonian, $\phi (k)$, with $R\phi (k)=\pm \phi (k)$. Here, $R$ is the reflection operator, and  $\phi(k)$ is $k$-independent because, by definition, a decomposable system lacks hopping between subsystems (unit cells).
\subsection{\texorpdfstring{The lattice models $H_{p,n}$}{\textmu}}
\begin{figure}[htb!]
\centering
\includegraphics[width=0.4\textwidth]{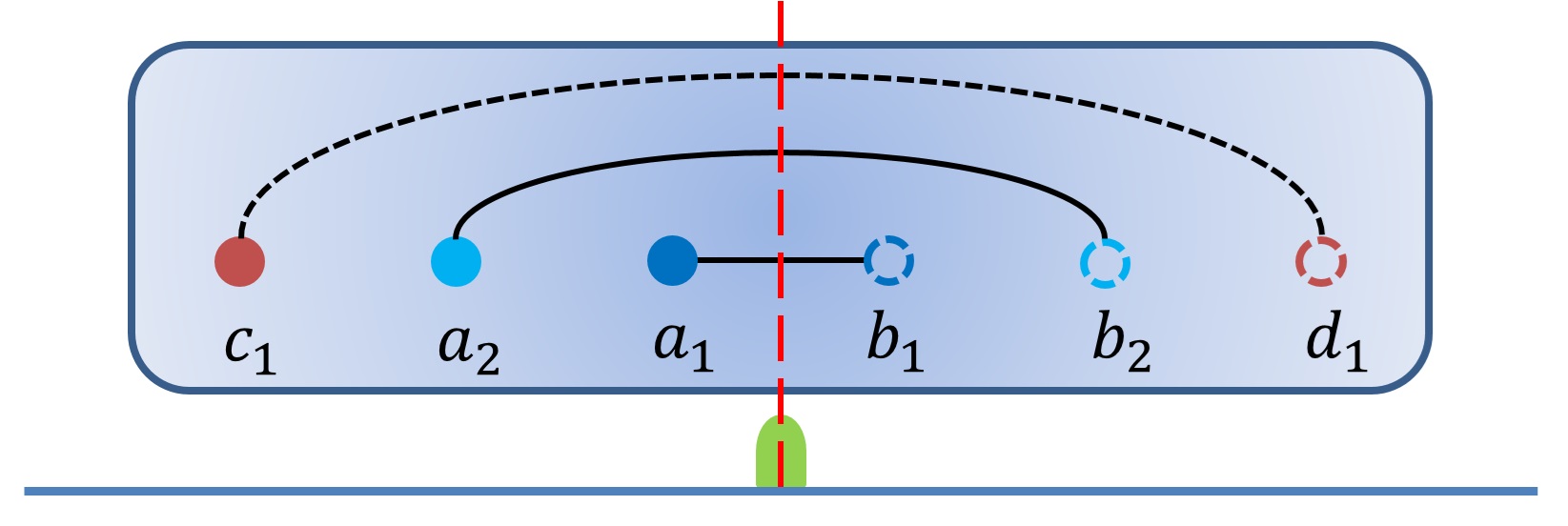}
\caption{Schematic depiction of a subsystem of $H_{2,1}$. The solid black line and the dashed black line represent positive and negative hopping strengths, respectively. The red dashed line marks the midpoint of this subsystem, where a charge, represented by a green wave packet, is located.}
\label{H21}
\end{figure}
Here, we introduce a series of lattice models $H_{p,n}$, where $p,n\in\mathbb{Z}_{\ge 0}$. These models are decomposable and are capable of capturing all classes of free-fermion and interacting crystalline SPT phases protected by $U(1)$ and reflection symmetry. The Hamiltonian of $H_{p,n}$ is
\bea
H_{p,n}=\sum_{j=1}^{N_{\text{tot}}} \left[\sum^{p}_{m} (v_m a^{\dag}_{m,j}b_{m,j}+h.c.)+\sum^{n}_{m^{'}} (w_{m^{'}} c^{\dag}_{m^{'},j}d_{m^{'},j}+h.c.)\right],
\eea
where $j$ indexes the subsystems, $N_{\text{tot}}$ represents the total number of subsystems comprising $H_{p,n}$, $v_m>0$, and $w_{m^{'}}<0$. These systems respect the reflection symmetry with $Ra^{\dag}_{m,j}R^{-1}=b^{\dag}_{m,N_{\text{tot}}-j+1}$ and $Rc^{\dag}_{m,j}R^{-1}=d^{\dag}_{m,N_{\text{tot}}-j+1}$. The model $H_{p,n}$ can be interpreted as a system composed of subsystems formed by stacking $p$ dipoles with positive hopping strengths and $n$ dipoles with negative hopping strengths. In Fig.~\ref{H21}, we present a schematic depiction of $H_{2,1}$.

The many-body ground state of $H_{p,n}$ at half-filling can be expressed as
\bea
\ket{\psi}=\prod_{j=1}^{N_{\text{tot}}}\left[\prod^{p}_{m}\alpha_{m}(a^{\dag}_{m,j}-b^{\dag}_{m,j})\prod^{n}_{m^{'}}\alpha_{m^{'}}(c^{\dag}_{m^{'},j}+d^{\dag}_{m^{'},j})\right]\ket{0},
\eea
where $\alpha_{m}$ and $\alpha_{m^{'}}$ are coefficients. For non-trivial cases, $N_{\text{tot}}$ must be odd, as the reflection center must be positioned at the midpoint of a subsystem. We use the index $j=C$ to mark the subsystem with its midpoint at the reflection center. As mentioned in Sec.~\ref{Decomposable systems and quantum numbers}, the quantum number $(N_C,R_C)_I$ can be determined by the many-body ground state of the subsystem with $j=C$
\bea
\ket{\psi}_C=\prod^{p}_{m}\alpha_{m}(a^{\dag}_{m,C}-b^{\dag}_{m,C})\prod^{n}_{m^{'}}\alpha_{m^{'}}(c^{\dag}_{m^{'},C}+d^{\dag}_{m^{'},C})\ket{0},
\eea
with $Ra^{\dag}_{m,C}R^{-1}=b^{\dag}_{m,C}$ and $Rc^{\dag}_{m,C}R^{-1}=d^{\dag}_{m,C}$. By the definition~\eqref{NCRC reflection def}, the interacting quantum number of $H_{p,n}$ is $(n+p,\text{$p$ mod 2})$, which determines the class $H_{p,n}$ falls into 
\bea\label{interacting class for Hpn}
[\text{$n-p$ mod 4}].
\eea

As stated in Sec.~\ref{Decomposable systems and quantum numbers}, if the reflection center is the same as the midpoint of the unit cell, the free-fermion quantum number $(n_{+},n_{-})_{f}$ can be determined by the occupied states of the corresponding Bloch Hamiltonian. To illustrate this, we first transform $H_{p,n}$ into the Bloch Hamiltonian $\mathscr{H}_{p,n}(k)$ through a Fourier transformation
\bea
\mathscr{H}_{p,n}(k)=
\begin{pmatrix}
  &   &   &   &   &   &   &   &   &   &   & w_1 \\
  &   &   &   &   &   &   &   &   &   & \iddots &   \\
  &   &   &   &   &   &   &   &   & w_n &   &   \\
  &   &   &   &   &   &   &   & v_1 &   &   &   \\
  &   &   &   &   &   &   & \iddots &   &   &   &   \\
  &   &   &   &   &   & v_p &   &   &   &   &   \\
  &   &   &   &   & v_p &   &   &   &   &   &   \\
  &   &   &   & \iddots &   &   &   &   &   &   &   \\
  &   &   & v_1 &   &   &   &   &   &   &   &   \\
  &   & w_n &   &   &   &   &   &   &   &   &   \\
  & \iddots &   &   &   &   &   &   &   &   &   &   \\
w_1 &   &   &   &   &   &   &   &   &   &   &   \\
\end{pmatrix}
,
\eea
where the chosen basis is $\{c_{1},\ldots,c_{n},a_{1},\ldots,a_{p},b_{1},\ldots,b_{p},d_{1},\ldots,d_{n}\}$. The matrix representation of $R$ acting on $\mathscr{H}_{p,n}(k)$ is an anti-diagonal matrix with all anti-diagonal entries equal to 1. When $N_{\text{tot}}$ is odd, the $0d$ SPT states in the 0-cell can be represented by the occupied states of $\mathscr{H}_{p,n}(k)$, denoted as $\phi(k)$. In half-filling cases, we have $\phi_{+,m}(k)$ and $\phi_{-,m^{'}}(k)$, where $\phi_{+,m}$ has non-zero components with $a_{m}=-b_{m}$, and $\phi_{-,m^{'}}$ has non-zero components with $c_{m^{'}}=d_{m^{'}}$. Since $Ra_{m}R^{-1}=b_{m}$ and $Rc_{m^{'}}R^{-1}=d_{m^{'}}$, it is evident that there are $p$ states with a reflection eigenvalue $R=-1$ and $n$ states with a reflection eigenvalue $R=1$. As a result, the free-fermion quantum number of $H_{p,n}$ is given by $(n,p)_f$, which indicates the class $H_{p,n}$ falls into
\bea\label{free class for Hpn}
[n-p].
\eea
By comparing eq.~\eqref{interacting class for Hpn} and eq.~\eqref{free class for Hpn}, one can see the validity of the connection between free-fermion and interacting classes~\eqref{reflection interaction effect}.

\end{appendices}
\bibliographystyle{unsrt}
\bibliography{citation}% Produces the bibliography via BibTeX.

\begin{thebibliography}{10}

\bibitem{IQHE}
D.~J. Thouless, M.~Kohmoto, M.~P. Nightingale, and M.~den Nijs.
\newblock Quantized hall conductance in a two-dimensional periodic potential.
\newblock {\em Phys. Rev. Lett.}, 49:405--408, Aug 1982.

\bibitem{IQSHE1}
C.~L. Kane and E.~J. Mele.
\newblock Quantum spin hall effect in graphene.
\newblock {\em Phys. Rev. Lett.}, 95:226801, Nov 2005.

\bibitem{IQSHE2}
C.~L. Kane and E.~J. Mele.
\newblock ${Z}_{2}$ topological order and the quantum spin hall effect.
\newblock {\em Phys. Rev. Lett.}, 95:146802, Sep 2005.

\bibitem{IQSHE3}
B.~Andrei Bernevig, Taylor~L. Hughes, and Shou-Cheng Zhang.
\newblock Quantum spin hall effect and topological phase transition in hgte quantum wells.
\newblock {\em Science}, 314(5806):1757–1761, December 2006.

\bibitem{IQSHE4}
Markus Konig, Steffen Wiedmann, Christoph Brune, Andreas Roth, Hartmut Buhmann, Laurens~W. Molenkamp, Xiao-Liang Qi, and Shou-Cheng Zhang.
\newblock Quantum spin hall insulator state in hgte quantum wells.
\newblock {\em Science}, 318(5851):766–770, November 2007.

\bibitem{TI1}
Liang Fu, C.~L. Kane, and E.~J. Mele.
\newblock Topological insulators in three dimensions.
\newblock {\em Phys. Rev. Lett.}, 98:106803, Mar 2007.

\bibitem{TI2}
M.~Z. Hasan and C.~L. Kane.
\newblock Colloquium: Topological insulators.
\newblock {\em Reviews of Modern Physics}, 82(4):3045–3067, November 2010.

\bibitem{AZclass}
Alexander Altland and Martin~R. Zirnbauer.
\newblock Nonstandard symmetry classes in mesoscopic normal-superconducting hybrid structures.
\newblock {\em Phys. Rev. B}, 55:1142--1161, Jan 1997.

\bibitem{Ktheory1}
Andreas~P. Schnyder, Shinsei Ryu, Akira Furusaki, and Andreas W.~W. Ludwig.
\newblock Classification of topological insulators and superconductors in three spatial dimensions.
\newblock {\em Phys. Rev. B}, 78:195125, Nov 2008.

\bibitem{Ktheory2}
Alexei Kitaev, Vladimir Lebedev, and Mikhail Feigel’man.
\newblock Periodic table for topological insulators and superconductors.
\newblock In {\em AIP Conference Proceedings}. AIP, 2009.

\bibitem{Ktheory3}
Shinsei Ryu, Andreas~P Schnyder, Akira Furusaki, and Andreas W~W Ludwig.
\newblock Topological insulators and superconductors: tenfold way and dimensional hierarchy.
\newblock {\em New Journal of Physics}, 12(6):065010, June 2010.

\bibitem{Ktheory4}
Ching-Kai Chiu, Jeffrey C.~Y. Teo, Andreas~P. Schnyder, and Shinsei Ryu.
\newblock Classification of topological quantum matter with symmetries.
\newblock {\em Rev. Mod. Phys.}, 88:035005, Aug 2016.

\bibitem{SPT1}
Zheng-Cheng Gu and Xiao-Gang Wen.
\newblock Tensor-entanglement-filtering renormalization approach and symmetry-protected topological order.
\newblock {\em Phys. Rev. B}, 80:155131, Oct 2009.

\bibitem{SPT2}
Xie Chen, Zheng-Cheng Gu, and Xiao-Gang Wen.
\newblock Local unitary transformation, long-range quantum entanglement, wave function renormalization, and topological order.
\newblock {\em Phys. Rev. B}, 82:155138, Oct 2010.

\bibitem{BBC1}
Shinsei Ryu and Yasuhiro Hatsugai.
\newblock Topological origin of zero-energy edge states in particle-hole symmetric systems.
\newblock {\em Phys. Rev. Lett.}, 89:077002, Jul 2002.

\bibitem{BBC2}
Hui Li and F.~D.~M. Haldane.
\newblock Entanglement spectrum as a generalization of entanglement entropy: Identification of topological order in non-abelian fractional quantum hall effect states.
\newblock {\em Phys. Rev. Lett.}, 101:010504, Jul 2008.

\bibitem{BBC3}
Y.~Hatsugai.
\newblock Bulk-edge correspondence in graphene with/without magnetic field: Chiral symmetry, dirac fermions and edge states.
\newblock {\em Solid State Communications}, 149(27):1061--1067, 2009.

\bibitem{BBC4}
Roger S.~K. Mong and Vasudha Shivamoggi.
\newblock Edge states and the bulk-boundary correspondence in dirac hamiltonians.
\newblock {\em Phys. Rev. B}, 83:125109, Mar 2011.

\bibitem{BBC5}
P.~Delplace, D.~Ullmo, and G.~Montambaux.
\newblock Zak phase and the existence of edge states in graphene.
\newblock {\em Phys. Rev. B}, 84:195452, Nov 2011.

\bibitem{BBC6}
Gian~Michele Graf and Marcello Porta.
\newblock Bulk-edge correspondence for two-dimensional topological insulators.
\newblock {\em Communications in Mathematical Physics}, 324(3):851–895, October 2013.

\bibitem{BBC7}
János~K. Asbóth, László Oroszlány, and András Pályi.
\newblock {\em A Short Course on Topological Insulators}.
\newblock Springer International Publishing, 2016.

\bibitem{BBC8}
Yang Peng, Yimu Bao, and Felix von Oppen.
\newblock Boundary green functions of topological insulators and superconductors.
\newblock {\em Phys. Rev. B}, 95:235143, Jun 2017.

\bibitem{BBC9}
Chen-Shen Lee, Iao-Fai Io, and Hsien-chung Kao.
\newblock Winding number and zak phase in multi-band ssh models.
\newblock {\em Chinese Journal of Physics}, 78:96–110, August 2022.

\bibitem{BBC10}
Chen-Shen Lee.
\newblock A linear algebra-based approach to understanding the relation between the winding number and zero-energy edge states.
\newblock {\em SciPost Phys. Core}, 7:003, 2024.

\bibitem{3DTIwithTR}
Xiao-Liang Qi, Taylor~L. Hughes, and Shou-Cheng Zhang.
\newblock Topological field theory of time-reversal invariant insulators.
\newblock {\em Phys. Rev. B}, 78:195424, Nov 2008.

\bibitem{edgedegeneracy1}
Lukasz Fidkowski and Alexei Kitaev.
\newblock Effects of interactions on the topological classification of free fermion systems.
\newblock {\em Phys. Rev. B}, 81:134509, Apr 2010.

\bibitem{edgedegeneracy2}
Evelyn Tang and Xiao-Gang Wen.
\newblock Interacting one-dimensional fermionic symmetry-protected topological phases.
\newblock {\em Phys. Rev. Lett.}, 109:096403, Aug 2012.

\bibitem{reductionSPT1}
Zheng-Cheng Gu and Michael Levin.
\newblock Effect of interactions on two-dimensional fermionic symmetry-protected topological phases with ${Z}_{2}$ symmetry.
\newblock {\em Phys. Rev. B}, 89:201113, May 2014.

\bibitem{reductionSPT2}
Yi-Zhuang You and Cenke Xu.
\newblock Symmetry-protected topological states of interacting fermions and bosons.
\newblock {\em Phys. Rev. B}, 90:245120, Dec 2014.

\bibitem{reductionSPT3}
Takahiro Morimoto, Akira Furusaki, and Christopher Mudry.
\newblock Breakdown of the topological classification $\mathbb{Z}$ for gapped phases of noninteracting fermions by quartic interactions.
\newblock {\em Phys. Rev. B}, 92:125104, Sep 2015.

\bibitem{TCI1}
Liang Fu and C.~L. Kane.
\newblock Topological insulators with inversion symmetry.
\newblock {\em Phys. Rev. B}, 76:045302, Jul 2007.

\bibitem{TCI2}
Liang Fu.
\newblock Topological crystalline insulators.
\newblock {\em Phys. Rev. Lett.}, 106:106802, Mar 2011.

\bibitem{TCI3}
Yoichi Ando and Liang Fu.
\newblock Topological crystalline insulators and topological superconductors: From concepts to materials.
\newblock {\em Annual Review of Condensed Matter Physics}, 6(1):361–381, March 2015.

\bibitem{TCI4}
Takahiro Morimoto and Akira Furusaki.
\newblock Topological classification with additional symmetries from clifford algebras.
\newblock {\em Phys. Rev. B}, 88:125129, Sep 2013.

\bibitem{TCI5}
Ching-Kai Chiu, Hong Yao, and Shinsei Ryu.
\newblock Classification of topological insulators and superconductors in the presence of reflection symmetry.
\newblock {\em Phys. Rev. B}, 88:075142, Aug 2013.

\bibitem{TCI6}
Ching-Kai Chiu and Andreas~P. Schnyder.
\newblock Classification of reflection-symmetry-protected topological semimetals and nodal superconductors.
\newblock {\em Phys. Rev. B}, 90:205136, Nov 2014.

\bibitem{TCI7}
Ken Shiozaki and Masatoshi Sato.
\newblock Topology of crystalline insulators and superconductors.
\newblock {\em Phys. Rev. B}, 90:165114, Oct 2014.

\bibitem{CSPT1}
Norbert Schuch, David P\'erez-Garc\'{\i}a, and Ignacio Cirac.
\newblock Classifying quantum phases using matrix product states and projected entangled pair states.
\newblock {\em Phys. Rev. B}, 84:165139, Oct 2011.

\bibitem{CSPT2}
Xie Chen, Zheng-Cheng Gu, and Xiao-Gang Wen.
\newblock Classification of gapped symmetric phases in one-dimensional spin systems.
\newblock {\em Phys. Rev. B}, 83:035107, Jan 2011.

\bibitem{CSPT3}
Xie Chen, Zheng-Cheng Gu, and Xiao-Gang Wen.
\newblock Complete classification of one-dimensional gapped quantum phases in interacting spin systems.
\newblock {\em Phys. Rev. B}, 84:235128, Dec 2011.

\bibitem{CSPT4}
Chang-Tse Hsieh, Olabode~Mayodele Sule, Gil~Young Cho, Shinsei Ryu, and Robert~G. Leigh.
\newblock Symmetry-protected topological phases, generalized laughlin argument, and orientifolds.
\newblock {\em Phys. Rev. B}, 90:165134, Oct 2014.

\bibitem{CSPT5}
Chang-Tse Hsieh, Takahiro Morimoto, and Shinsei Ryu.
\newblock Cpt theorem and classification of topological insulators and superconductors.
\newblock {\em Phys. Rev. B}, 90:245111, Dec 2014.

\bibitem{CSPT6}
Yohei Fuji, Frank Pollmann, and Masaki Oshikawa.
\newblock Distinct trivial phases protected by a point-group symmetry in quantum spin chains.
\newblock {\em Phys. Rev. Lett.}, 114:177204, May 2015.

\bibitem{CSPT7}
Gil~Young Cho, Chang-Tse Hsieh, Takahiro Morimoto, and Shinsei Ryu.
\newblock Topological phases protected by reflection symmetry and cross-cap states.
\newblock {\em Phys. Rev. B}, 91:195142, May 2015.

\bibitem{free-to-interactingmap}
Naren Manjunath, Vladimir Calvera, and Maissam Barkeshli.
\newblock Characterization and classification of interacting ($2+1$)-dimensional topological crystalline insulators with orientation-preserving wallpaper groups.
\newblock {\em Phys. Rev. B}, 109:035168, Jan 2024.

\bibitem{AHSSsource}
Michael~F Atiyah and Friedrich Hirzebruch.
\newblock {\em Vector bundles and homogeneous spaces}, page 196–222.
\newblock London Mathematical Society Lecture Note Series. Cambridge University Press, 1972.

\bibitem{constructionCSPTfromlowerd1}
Hao Song, Sheng-Jie Huang, Liang Fu, and Michael Hermele.
\newblock Topological phases protected by point group symmetry.
\newblock {\em Phys. Rev. X}, 7:011020, Feb 2017.

\bibitem{constructionCSPTfromlowerd2}
Sheng-Jie Huang, Hao Song, Yi-Ping Huang, and Michael Hermele.
\newblock Building crystalline topological phases from lower-dimensional states.
\newblock {\em Phys. Rev. B}, 96:205106, Nov 2017.

\bibitem{Lower-dconstruction1}
Zhida Song, Chen Fang, and Yang Qi.
\newblock Real-space recipes for general topological crystalline states.
\newblock {\em Nature Communications}, 11(1), August 2020.

\bibitem{Lower-dconstruction2}
Meng Cheng and Chenjie Wang.
\newblock Rotation symmetry-protected topological phases of fermions.
\newblock {\em Phys. Rev. B}, 105:195154, May 2022.

\bibitem{Lower-dconstruction3}
Jian-Hao Zhang, Shuo Yang, Yang Qi, and Zheng-Cheng Gu.
\newblock Real-space construction of crystalline topological superconductors and insulators in 2d interacting fermionic systems.
\newblock {\em Phys. Rev. Res.}, 4:033081, Jul 2022.

\bibitem{2024realspaceconstruction}
Jian-Hao {Zhang}, Shang-Qiang {Ning}, Yang {Qi}, and Zheng-Cheng {Gu}.
\newblock {Construction and classification of crystalline topological superconductor and insulators in three-dimensional interacting fermion systems}.
\newblock {\em arXiv e-prints}, page arXiv:2204.13558, April 2022.

\bibitem{interactingAHSS}
Ken Shiozaki, Charles~Zhaoxi Xiong, and Kiyonori Gomi.
\newblock Generalized homology and atiyah–hirzebruch spectral sequence in crystalline symmetry protected topological phenomena.
\newblock {\em Progress of Theoretical and Experimental Physics}, 2023(8), July 2023.

\bibitem{freeAHSSinkspace}
Ken Shiozaki, Masatoshi Sato, and Kiyonori Gomi.
\newblock Atiyah-hirzebruch spectral sequence in band topology: General formalism and topological invariants for 230 space groups.
\newblock {\em Phys. Rev. B}, 106:165103, Oct 2022.

\bibitem{freeAHSSinrealspace}
Nobuyuki Okuma, Masatoshi Sato, and Ken Shiozaki.
\newblock Topological classification under nonmagnetic and magnetic point group symmetry: Application of real-space atiyah-hirzebruch spectral sequence to higher-order topology.
\newblock {\em Phys. Rev. B}, 99:085127, Feb 2019.

\bibitem{TKtheory1}
Daniel~S. Freed and Gregory~W. Moore.
\newblock {Twisted Equivariant Matter}.
\newblock {\em Annales Henri Poincaré}, 12 2013.

\bibitem{TKtheory2}
Guo~Chuan Thiang.
\newblock {On the K-Theoretic Classification of Topological Phases of Matter}.
\newblock {\em Annales Henri Poincaré}, 04 2016.

\bibitem{BIQHE}
Yuan-Ming Lu and Ashvin Vishwanath.
\newblock Theory and classification of interacting integer topological phases in two dimensions: A chern-simons approach.
\newblock {\em Phys. Rev. B}, 86:125119, Sep 2012.

\bibitem{Ourcrystallineequivalence}
Chen-Shen Lee, Ken Shiozaki, and Chang-Tse Hsieh.
\newblock {Crystalline-equivalent topological phases of many-body fermionic systems in 1+1 dimensions}.
\newblock {\em arXiv e-prints}, 2024.

\end{thebibliography}

\end{document}